\documentclass[superscriptaddress,twocolumn,10pt,pra,aps,showpacs,longbibliography]{revtex4-1}
\usepackage{float}
\floatstyle{boxed}
\usepackage{wrapfig}
\usepackage{array}
\usepackage{graphicx}
\usepackage{amsbsy}
\usepackage[utf8x]{inputenc}
\usepackage{epstopdf}
\usepackage{amsmath,amssymb,amsfonts}
\usepackage{dsfont}
\usepackage{cprotect}

\usepackage[dvipsnames]{xcolor}

\def \am{\hat a}
\def \ap{\hat a^{\dagger}}

\newcommand{\bra}[1]{\langle #1|}
\newcommand{\ket}[1]{|#1\rangle}

\newcommand{\expec}[1]{\left\langle #1 \right\rangle}

\renewcommand{\eqref}[1]{\mbox{Eq.~(\ref{#1})}}
\renewcommand{\Re}[1]{{\rm Re}\left[#1 \right]}

\newcommand{\be}{\begin{equation}}
\newcommand{\ee}{\end{equation}}
\newcommand{\bea}{\begin{eqnarray}}
\newcommand{\eea}{\end{eqnarray}}

\usepackage{url}
\usepackage[colorlinks]{hyperref}
\hypersetup{
	plainpages=true,
	breaklinks=true,
	hypertexnames=false,
	pageanchor=true,
	colorlinks=true,
	linkcolor={blue},
	citecolor={red},
	urlcolor={blue},
	anchorcolor={black}
}

\newcommand{\LL}{\mathcal{L}}
\newcommand{\DD}{\mathcal{D}}
\newcommand{\rhot}{\hat{\rho}(t)}
\newcommand{\de}{{\rm d}}
\newcommand{\sss}{\hat{\rho}_{\rm ss}}
\newcommand{\eig}[1]{\hat{\rho}_{#1}}

\makeatletter
\newcommand*\bigcdot{\mathpalette\bigcdot@{.5}}
\newcommand*\bigcdot@[2]{\mathbin{\vcenter{\hbox{\scalebox{#2}{$\m@th#1\bullet$}}}}}
\makeatother

\usepackage[normalem]{ulem} 

\begin{document}

	\author{Fabrizio Minganti }
	\email{fabrizio.minganti@riken.jp} 
	\affiliation{Theoretical
		Quantum Physics Laboratory, RIKEN Cluster for Pioneering Research,
		Wako-shi, Saitama 351-0198, Japan}
	\author{Ievgen I. Arkhipov}
	\email{ievgen.arkhipov@upol.cz} 
	\affiliation{Joint Laboratory of
		Optics of Palack\'y University and Institute of Physics of CAS,
		Faculty of Science, Palack\'y University, 17. listopadu 12, 771 46
		Olomouc, Czech Republic}
	\author{Adam Miranowicz}
	\email{miran@amu.edu.pl} 
	\affiliation{Theoretical Quantum Physics Laboratory, RIKEN Cluster
		for Pioneering Research, Wako-shi, Saitama 351-0198, Japan}
	\affiliation{Faculty of Physics, Adam
		Mickiewicz University, 61-614 Pozna\`n, Poland}
	\author{Franco Nori}
	\email{fnori@riken.jp}
	\affiliation{Theoretical Quantum Physics
		Laboratory, RIKEN Cluster for Pioneering Research, Wako-shi,
		Saitama 351-0198, Japan} \affiliation{Physics Department, The
		University of Michigan, Ann Arbor, Michigan 48109-1040, USA}

	\title{Correspondence between dissipative phase transitions of light and time
		crystals}
	
	\begin{abstract}

		We predict the emergence of a time crystal generated by an incoherently driven and dissipative nonlinear optical oscillator, where the nonlinearity also comes from dissipation. We show that a second-order dissipative phase transition of light occurs in the frame rotating at the cavity frequency, while a boundary (dissipative) time crystal emerges in the laboratory frame. We relate these two phenomena
		by using the Liouvillian superoperator associated to the Lindblad
		master equation and its symmetries.  These results connect
		the emergence of a second-order dissipative phase transition and a dissipative time
		crystal in the thermodynamic limit, allowing to interpret them as
		the same phenomenon in terms of the Liouvillian spectrum, but
		just in different frames.
		
	\end{abstract}
	
	\date{\today}
	
	\maketitle


\section{Introduction} 

Correspondences play a central role in physics. Seminal examples include the bulk-edge correspondence, linking the presence of edge states to bulk topological invariants~\cite{Kane2005}, and the AdS/CFT (gauge/gravity) correspondence, which establishes a duality between conformal field theories and supergravity~\cite{Maldacena1999}.
	Correspondences characterize related but yet different features of systems by linking their seemingly different properties.
	This article discusses the correspondence of two nonanalytical  phenomena in open many-body quantum physics: {\it dissipative} phase transitions (DPTs) \cite{CarmichaelPRX15,FinkNatPhys18,MingantPRA18_Spectral} and {\it boundary} \footnote{The name boundary time crystals is used in Refs.~\cite{IeminiPRL18, LLedoNJP20} as a synonym of dissipative time crystal. Indeed, a dissipative system can be seen as the boundary of the universe (where the universe is the sum of the system and its environment). Tracing out all the degrees of freedom of the environment, the boundary time crystals emerge.} time crystals (BTCs)\cite{IeminiPRL18, LLedoNJP20}.
	
	The driven-dissipative physics of light is the focus of intense research, fostered by the achievements of non-negligible photon-photon interactions and sizeable light-matter couplings ~\cite{Carusotto_RMP_2013_quantum_fluids_light,kockum2019}. Nonlinear resonators can be realized, e.g., in semiconductor microcavities~\cite{DelteilNatMat19} and superconducting circuits~\cite{YouNat11,Gu2017,Kjaergaard20}.
	These systems are driven out of their thermal equilibrium and do not obey the paradigms of thermodynamics \cite{BreuerBookOpen,LidarLectureNotes}.
	Even in the absence of free-energy analysis, open quantum systems display critical phenomena, such as DPTs \cite{MingantPRA18_Spectral}. 
	These out-of-equilibrium phases were initially
	studied in connection with lasing-like phenomena \cite{DeGiorgio1970,BonifacioPRA78,Mollow67,Drummond1978}.
	More recently, several examples of DPTs have been predicted for nonlinear optical systems \cite{WeimerPRL2015,MendozaPRA16,CasteelsPRA16,BartoloPRA16,Foss-FeigPRA17,BiellaPRA17,SavonaPRA17} and spin systems \cite{LeePRL13,CarmichaelPRX15,JinPRX16,RotaPRB17,HwangPRA18,RotaNJP18,RotaPRL19,HuybrechtsPRB20,GarbePRL20, LandaPRL20,CurtisarXiv20}.
	Experimentally, DPTs have been observed in lattices of superconducting resonators \cite{FitzpatrickPRX17}, Rydberg atoms in optical lattices \cite{Mueller_2012,BernienNAT2017}, optomechanical systems \cite{AspelmeyerRMP14,GilSantosPRL17}, exciton-polariton condensates \cite{KasprzakNAT2006,Carusotto_RMP_2013_quantum_fluids_light}, single superconducting cavities \cite{FinkPRX17}, and semiconductor micropillars \cite{RodriguezPRL17,FinkNatPhys18}.
	
	While a DPT is associated with the emergence of multiple steady states, a BTC is formed when permanent oscillations arise spontaneously in an otherwise time-translation invariant system \cite{IeminiPRL18,ShammahPRA18,TuckerNJP18,LLedoPRB19,TuckerNJP18,BucaNat19,SeiboldPRA20} (we will \emph{not} consider discrete time crystals in time-dependent systems, where oscillations develop at times which are multiples of the driving one \cite{SachaRPP17, GongPRL18, RieraQuantum2020}).
	Questions concerning the resilience of BTCs in extended lattice geometries including generalized noise have been raised~\cite{TuckerNJP18,ShammahPRA18}, and the connection between the crystalline phase and the symmetries of the related model has been partially explored \cite{GongPRL18,GambettaPRL19,LLedoPRB19, BucaNat19,ScarlatellaPRB19,LLedoNJP20}.
	
	In many of the previous works, the physics is captured by a Lindblad master equation.
	DPTs and BTCs can be understood as critical spectral properties of the Liouvillian superoperator, i.e., the matrix describing the evolution of the density matrix written in its vectorized form.
	A DPT emerges when one (or more) Liouvillian eigenvalues become zero in both real and imaginary parts \cite{KesslerPRA12,MingantPRA18_Spectral}.
	Similarly, for BTCs, the Liouvillian acquires eigenvalues with zero real part but nonzero imaginary one \cite{IeminiPRL18,ShammahPRA18,Bookerarxiv20}.
	
	In this article, we study the correspondence between DPTs and BTCs by investigating a single-mode cavity with incoherent drive and incoherent nonlinear dissipative processes \cite{BiellaPRA17,LebreuillyPRA17,TakemuraArXiv19}. 
	Using a simple change-of-reference transformation, we show that in the frame rotating at the cavity frequency (say, the R-frame) the system displays typical features of a DPT, while in the laboratory frame (the L-frame) the system undergoes time crystallization. Note that the Liouvillian is time-independent in both frames.
	Hence, we prove that for this highly-symmetrical model, a \emph{second-order DPT and a BTC are one and the same phenomenon but in two different reference frames}. We stress that this fact can be seen as a proof of the existence of BTCs in open systems.
	
	The article is structured as follows.
	In Sec.~\ref{Sec:Model} we introduce a single-mode cavity with incoherent drive and incoherent nonlinear dissipative processes. We detail its master equation and the associated Liouvillian superopertator in the L-frame. We discuss the L-frame Liouvillian properties and we show how to obtain the R-frame Liouvillian. We also briefly recall  the main properties of time crystals and dissipative phase transitions.
	In Sec.~\ref{Sec:Model1} we numerically demonstrate the correspondence between BTCs and DPTs for the model under consideration. In Sec.~\ref{Sec:Model2} we show the same correspondence but in a different nonlinear system, i.e., the Scully-Lamb laser model. We derive our conclusion in Sec.~\ref{Sec:Conclusion}.
	
	\section{The model}\label{Sec:Model}
Hereafter, we consider the Hamiltonian ${\hat{H}=\omega_c \hat{a}^\dagger \hat{a}}$ of a harmonic oscillator (e.g., that of an optical cavity), where
	$\hat{a}$ ($\hat{a}^\dagger$) is the bosonic annihilation (creation) operator.
	This system interacts with a weak Markovian environment, and its state is captured by a density matrix $\rhot$ evolving under a Lindblad master equation ($\hbar=1$)
	\begin{equation}\label{Lindblad1}
	\frac{\de}{\de t}\rhot={\cal L}\hat\rho(t)=-i\left[\hat H,\rhot\right] +\sum\limits_{j} \DD[\hat{L}_j]\rhot.
	\end{equation}
	Here, $\LL$ is the Liouvillian superoperator \cite{LidarLectureNotes,BreuerBookOpen}, while $\DD[\hat{L}_j]$ are the so-called Lindblad dissipators, whose action is 
	\begin{equation}
	\DD[\hat{L}_j]\rhot=\hat L_i\rhot\hat L_j^{\dagger} - \frac{\hat L_j^{\dagger}\hat L_j\rhot+\rhot\hat L_j^{\dagger}\hat L_j}{2}.
	\end{equation}
	The operators $\hat{L}_j$ are called jump operators and they describe how the environment acts on the system inducing loss and gain of bosons, energy, and information.
	In the following, we consider 
	\begin{equation}\label{L123}  
	\begin{split}
	&\hat L_1 = \sqrt{\Gamma} \am, \ \, \hat L_2 = \sqrt{\eta}\,\am^2, \ \, \hat L_3 = \sqrt{\xi}\,\ap,
	\end{split}
	\end{equation}
	where $\hat{L}_1$ represents the single-particle loss, $\hat{L}_2$ the simultaneous loss of two photons, and $\hat{L}_3$ the incoherent drive (gain).
	The parameter $\Gamma$ is the inverse of the photon lifetime, $\xi$ represents the medium gain (incoherent drive) strength, and $\eta$ is the two-photon dissipation rate. 
	Such a model can be realized, e.g., considering an engineered thermal drive \cite{BiellaPRA17,LebreuillyPRA17} in the presence of an engineered two-photon dissipation \cite{LeghtasScience15,LescanneNatPhys2020}.
	
	The system exhibits a $U(1)$ symmetry since any transformation $\hat{a}\to \hat{a}\exp({i \phi})$ for an arbitrary real number $\phi$ leaves \eqref{Lindblad1} unchanged  (see Appendix~\ref{App:Properties}).
	While Noether's theorem implies the presence of a conserved quantity for $U(1)$ Hamiltonian symmetries, this is not the case for the Liouvillian under consideration. Indeed, $U(1)$ is a weak symmetry of \eqref{Lindblad1} and the extension of Noether's theorem for open systems require a strong symmetry \cite{BaumgartnerJPA08,BucaNPJ2012,AlbertPRA14}. Nonetheless, a Liouvillian symmetry constrains the dynamics of the system and influence its DPT \cite{MingantPRA18_Spectral}.

\subsection{Liouvillian spectrum}
	A central role in the following discussion is played by the steady state $\sss$, i.e., 
	the matrix which is stationary
	under the action of the Liouvillian: $\partial_t \sss=\LL \sss = 0$.
	For the system under consideration, if $\eta>0$ there is a unique steady state and, therefore, $\sss=\lim_{t \to \infty} \rhot$.
	The dynamics of the system cannot be obtained from the steady state alone. Having introduced the Liouvillian $\LL$, we define its eigenvalues $\lambda_i$ (representing typical decay times) and eigenmatrices $\eig{i}$ (encoding the states explored along the dynamics) via
	\begin{equation}
	\LL \eig{i}=\lambda_i \eig{i}.
	\end{equation}
	We order the eigenvalues $\lambda_i$ in such a way that $|\Re{\lambda_0}|<|\Re{\lambda_1}| <\dots <|\Re{\lambda_n}| < \dots$, i.e., the eigenvalues are ordered by their real part.
	In this regard, the steady state is proportional to $\eig{0}$, i.e., the eigenmatrix of the Liouvillian associated to  $\lambda_0=0$.
	In the following, we will use QuTiP \cite{qutip1,qutip2} to numerically diagonalize the Liouvillian and obtain its eigenvalues and eigenmatrices.

\subsection{Rotating frame of reference}
	The Hamiltonian dependence of the Liouvillian is eliminated in the frame which rotates at the cavity frequency $\omega_c$.
	The density matrix in the R-frame becomes 
	\begin{equation}
	\hat{\rho}^R(t) = \exp\left(i \omega_c t \hat{a}^\dagger \hat{a}\right) \rhot  \exp\left(-i \omega_c t \hat{a}^\dagger \hat{a}\right).
	\end{equation}
	Since all the dissipators in Eq.~(\ref{L123}) are $U(1)$-symmetric, they are unchanged by the transformation. Hence, the Liouvillian $\LL^R$ in the R-frame, defined  by $\partial_t \hat{\rho}^{R}(t)=\LL^R \hat{\rho}^{R}(t)$, is simply 
	\begin{equation}\label{Eq:Rotating_Liouvillian}
	\LL^R\hat{\rho}^{R}(t) = \sum_{j=1}^{3} \DD[\hat{L}_j]\hat{\rho}^{R}(t), \quad \LL^R \eig{i}^{R}=\lambda_i^{R} \eig{i}^{R},
	\end{equation}
	where $\lambda_i^{R}$ and $\eig{i}^{R}$ are its eigenvalues and eigenvectors, respectively.
	We stress two facts: (i) that both $\LL^R$ and $\LL$ are time independent; and (ii) the role of the symmetry in obtaining \eqref{Eq:Rotating_Liouvillian}.
	
	Normally, there is no trivial correspondence 
	between $\lambda_i^{R}$ and $\lambda_i$, or $\eig{i}$ and $\eig{i}^{R}$, due to the presence of ``centrifugal forces'' in the non-inertial R-frame. However, for the model under consideration one can explicitly compute $\LL \eig{i}^R$, and using the presence of a $U(1)$ symmetry demonstrate that there exist the following fundamental relations (the proof is provided in Appendix~\ref{App:Proof})
	\begin{equation}\label{Eq:Fundamental_correspondence}
	\eig{i}= \eig{i}^{R}\, , \quad \lambda_i = \lambda_i^{R} - i \omega_c k,
	\end{equation}
	where $k$ is an integer.

\subsection{Dissipative phase transitions and Boundary time crystals}
	A DPT is a discontinuous change in the steady-state density matrix $\sss$ of a time-independent open quantum system as a function of a parameter $\xi$, i.e., at the critical point $\xi_c$, $\partial_\xi^n \sss(\xi_c)=\infty$ for some integer $n$.
	Similarly to a closed quantum system, where the Hamiltonian energy gap vanishes at the critical point of a quantum phase transition, in a DPT the Liouvillian gap closes \cite{MingantPRA18_Spectral,LLedoPRB19}. Symmetries are pivotal in understanding second-order (i.e., $n=2$) DPTs \cite{MingantPRA18_Spectral,CurtisarXiv20}.
	For a second-order DPT triggered by a spontaneous symmetry-breaking, several eigenvalues close simultaneously (one for each generator of the symmetry group except the identity). For example, the DPT associated to a $Z_N$ symmetry breaking requires $\lambda_1=\lambda_2=\dots=\lambda_{N-1}=0$ \cite{MingantPRA18_Spectral}.
	This corresponds to the emergence of $N$ symmetry-breaking steady-state density matrices. For the $U(1)$ model under consideration, in the case of spontaneous symmetry breaking we expect infinitely-many eigenvalues to become zero.

	Similarly, we can interpret BTCs as a critical phenomenon. 
	Namely, given a time-independent Liouvillian $\LL$ which admits a unique steady state, a BTC appears in the thermodynamic limit if a characteristic frequency $\Omega$ appears such that the lowest-lying Liouvillian eigenvalues become $\lambda_{1, \, 2} = \pm i \Omega$.
	In the same way in which a crystal breaks the spatial translation invariance, $\LL$ breaks the time-translation invariance at $t \to \infty$ and it develops everlasting oscillations for an observable whose time dynamics is ruled by $\lambda_{1,2}$.
	There are several definitions of time crystals for closed ~\cite{Watanabe2015,SachaRPP17}, coherently driven~\cite{Sacha2015,SachaRPP17,Choi2017,Zhang2017}, and incoherently driven systems~\cite{TuckerNJP18}. We focus on the diagnostic of BTCs based on the Liouvillian gap~\cite{IeminiPRL18}.

\section{Numerical study}	
\label{Sec:Model1}
	
\subsection{Thermodynamic limit}
	Theoretically, critical phenomena can only emerge in the thermodynamic limit of an infinite number of particles. For extended systems (e.g., a lattice with $L$ sites), the thermodynamic limit can be obtained by increasing to infinity the system size ($L \to \infty$). 
	In a single bosonic resonator, we can use its infinite-dimensional Hilbert space to introduce the infinite number of photons necessary for criticality by an appropriate scaling of the parameters (see Refs.~\cite{CasteelsPRA17,BartoloPRA16,MingantPRA18_Spectral,HwangPRA18,GarbePRL20,FelicettiPRL20}). 
	Here, we can do it by introducing an effective parameter $N$, under which the number of photons transforms as $n \to N n$.
	Accordingly, $\{\Gamma, \xi, \eta\} \to \{\Gamma,  \xi, \eta/N\} $.
	One can interpret $N$ as the number of resonators in a Bose-Hubbard lattice, where each site is subject to the dissipators $\hat{L}_1$ and $\hat{L}_2$, but only the uniform Fourier mode $k=0$ is subject to the incoherent drive $\hat{L}_3$. In this case, \eqref{Lindblad1} describes the dynamics of the uniform mode $k=0$ of the lattice (see Appendix \ref{App:Thermodynamic_limit}).
	The dimensionless parameter $N$ induces a scale transformation that keeps unchanged the typical photon lifetime $1/\Gamma$. Thus, we use $\Gamma$ as a unit to study this problem.

	\begin{figure}
		\centering
		\includegraphics[width=0.95 \linewidth]{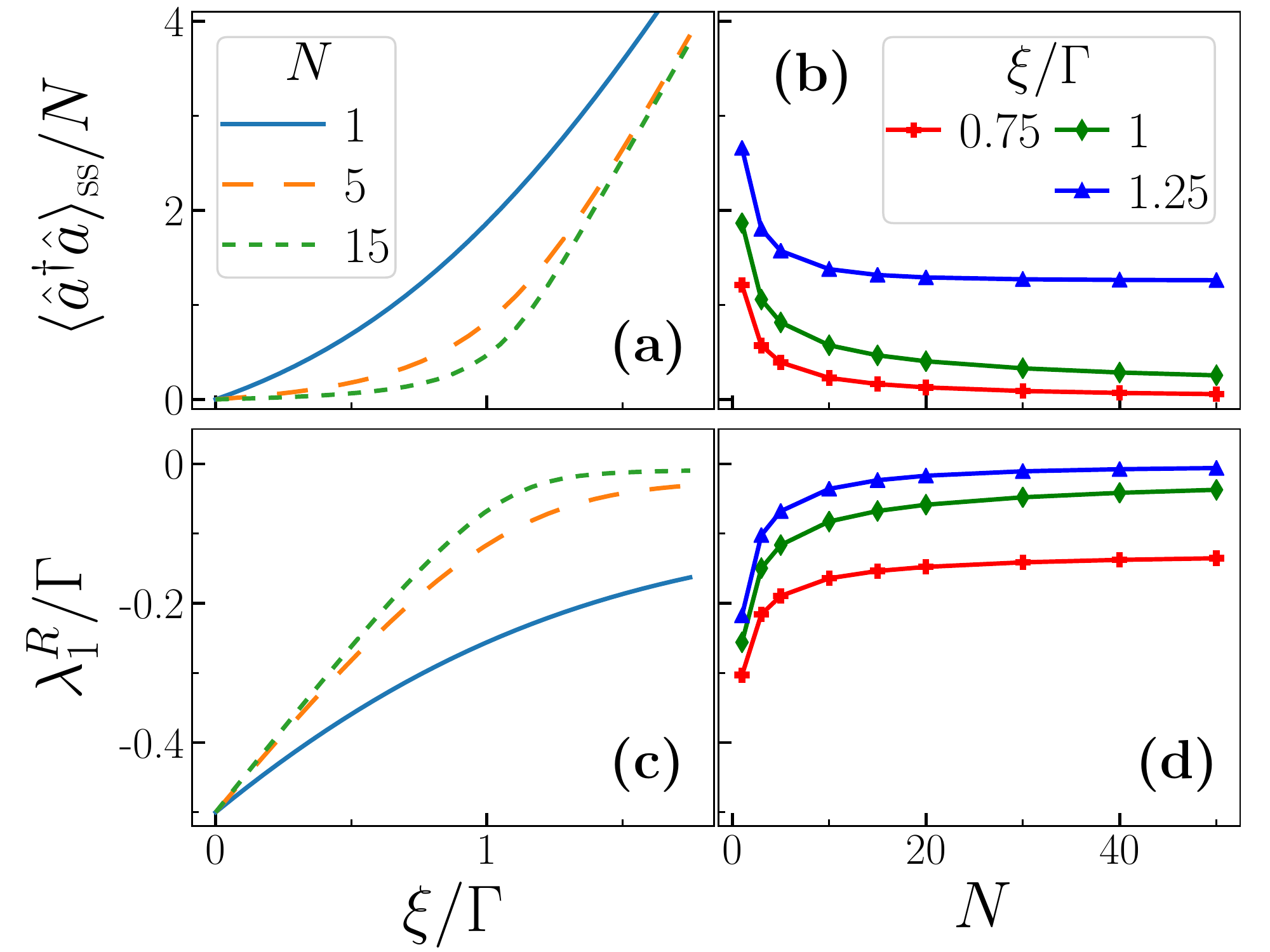}
		\caption{Onset of the DPT in the R-frame for $\eta/\Gamma=1$.
			(a) Steady-state rescaled number of photons $\expec{\hat{a}^\dagger \hat{a}}_{\rm ss}/N$ versus the incoherent drive strength $\xi/\Gamma$ for different values of the rescaling parameter $N$ (the thermodynamic limit is $N\to \infty$).
			(b) $\expec{\hat{a}^\dagger \hat{a}}_{\rm ss}/N$ as a function of $N$ for different $\xi$: $\xi/\Gamma=0.75$ (before the DPT), $\xi/\Gamma=1$ (critical point), and $\xi/\Gamma=1.25$ (after the DPT).    
			(c) Liouvillian gap $\lambda_1^R$ (in units of the damping rate $\Gamma$) versus  $\xi/\Gamma$ for different values of $N$.
			(d) $\lambda_1^R$ as a function of $N$.
		}
		\label{fig:phase_transition}
	\end{figure}
	
\subsection{Phase transition in the rotating frame}
\begin{figure}
		\centering
		\includegraphics[width=\linewidth]{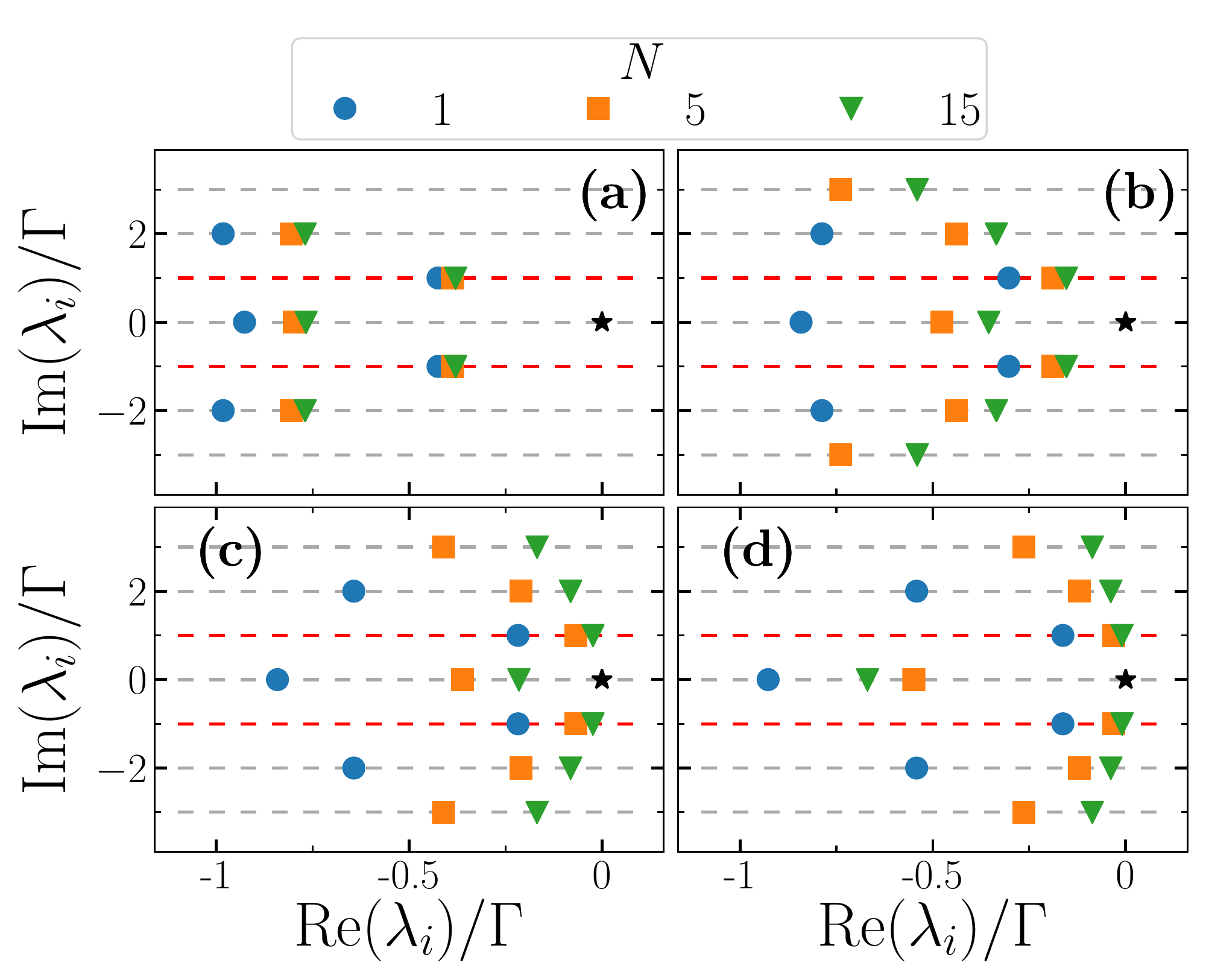}
		\caption{Real and imaginary parts of the Liouvillian eigenvalues in the L-frame for $\omega_c/\Gamma=1$, $\eta/\Gamma=0.1$, and $\xi/\Gamma$ equal to: (a) $0.25$, (b) $0.75$, (c) $1.25$, (d) $1.75$. Different markers represent different values of $N$, while the black stars represent the steady-state eigenvalue $\lambda_0=0$. The gray horizontal dashed lines represent $\operatorname{Im}(\lambda_i)=k \omega_c$, where $k$ is an integer. 
			The red horizontal dashed lines indicate those eigenvalues associated to the Liouvillian gap.
			The choice $\omega_c/\Gamma=1$ is arbitrary, and simply determines the imaginary part of $\lambda_i$.}
		\label{fig:time_crystal}
	\end{figure}

	In Fig.~\ref{fig:phase_transition}, we show the onset of the transition for the R-frame Liouvillian $\LL^R$ (the data are obtained by an exact diagonalization of $\LL^R$).
	As we see from Fig.~\ref{fig:phase_transition}(a), the change in the steady-state rescaled number of photons (i.e., $\expec{\hat{a}^\dagger \hat{a}}_{
		\rm ss}/N$) becomes more abrupt as we increase the value of the parameter $N$. In Fig.~\ref{fig:phase_transition}(b) we demonstrate that, by increasing $N$, $\expec{\hat{a}^\dagger \hat{a}}_{
		\rm ss}/N$ flows to zero before the phase transition, and converge to a finite value after it.
	In Fig.~\ref{fig:phase_transition}(c), instead, we plot the Liouvillian gap in the rotating frame $\lambda_1^R$, showing that the sharper the change in the photon number, the smaller is the Liouvillian gap.
	In Fig.~\ref{fig:phase_transition}(d) we show that by increasing $N$ the gap $\lambda_1^R$ is nonzero before the phase transition, and it become zero after the critical point.
	We have also verified that, in each symmetry sector, there is an eigenvalue closing at the DPT [it can also be concluded from Fig.~\ref{fig:time_crystal} using \eqref{Eq:Fundamental_correspondence}].
	Thus, the symmetry breaking is described by the simultaneous closure (in both real and imaginary parts) of infinitely many eigenvalues, one for each symmetry generator sector \cite{MingantPRA18_Spectral,RotaNJP18,JinPRB18,HuybrechtsPRB20}.

\subsection{Time crystal in the laboratory frame}
	Whereas in the R-frame the Liouvillian gap vanishes both in its real and imaginary parts ($\lambda_1^{R}\to0$), in the L-frame the imaginary part of $\lambda_1$ is never zero, according to Eq.~(\ref{Eq:Fundamental_correspondence}). That is, \emph{the DPT in the R-frame implies a BTC in the L-frame}.
	The BTCs are, therefore, a critical phenomenon appearing only in the thermodynamic limit.
	Thus, we diagonalize the Liouvillian in the L-frame for increasing values of $N$ for $\omega_c=\Gamma$.
	In Fig.~\ref{fig:time_crystal} we plot the real and imaginary parts of the low-lying part of the spectrum for different values of $\xi$ and $N$. As we enter the ``broken-symmetry phase'' and we increase $N$, there are slower and slower timescales, which are characterized by the imaginary parts of $\lambda_i$ being multiples integer of $\omega_c$, thus confirming \eqref{Eq:Fundamental_correspondence}.
	It follows that the field $\expec{\hat a}$ (and, thus, the electric field $\langle{\hat E}\rangle$) acquires an oscillatory behavior in the steady state in the thermodynamic limit.
	This can be interpreted as the emergence of a BTC \cite{IeminiPRL18}.
	Notice that this choice of $\omega_c$ is arbitrary and different values of $\omega_c$ would have produced time crystals characterised by a different frequency.
	
	Time crystallization is also accompanied by a discontinuity in the photon number in $\sss$ [as given in \eqref{Eq:Fundamental_correspondence}, all the results for the steady state are identical in the two frames]. Indeed, $\expec{\hat{a}^\dagger \hat{a}}_{\rm ss}$ in the L-frame is identical to that for the R-frame in Fig.~\ref{fig:phase_transition}(a).
	As such, the eigenvalue at $\operatorname{Im}(\lambda_i)=0$ plays a fundamental role: it is the one responsible for the non-analytical change of $\sss$ in the L-frame \cite{MingantPRA18_Spectral}.

\subsection{Two-time correlation functions}
	The model under consideration displays a second-order DPT in the R-frame associated to the spontaneous symmetry breaking of $U(1)$ induced by the interplay of the nonlinear dissipation $\hat{L}_2$ and the incoherent  drive $\hat{L}_3$.  
	However, in the L-frame a  BTC manifests itself, by breaking both $U(1)$ and time-translation symmetries, similarly to those in Ref.~\cite{ScarlatellaPRB19}. Indeed, in the L-frame, some of the eigenvalues $\lambda_i^R$ acquire an imaginary term proportional to $ik\omega_c$, according to Eq.~(\ref{Eq:Fundamental_correspondence}).
	That is, \eqref{Eq:Fundamental_correspondence} means that \emph{the onset of a DPT in the R-frame corresponds to a BTC in the L-frame}.
	
	We also stress that our model does not have a strong symmetry \cite{AlbertPRA14}, even in the thermodynamic limit. Furthermore, there is no Hamiltonian coherent process taking place in a well-defined ``decoherence-free subspace'' \cite{AlbertPRA14,AlbertPRX16, BucaNat19,Bookerarxiv20}. The emerging oscillations are the consequence of the same critical phenomenon leading to the $U(1)$ symmetry breaking, since both $\LL$ and $\LL^{R}$ are time-independent superoperators. This is the very same symmetry which allows to pass from one frame to the other without introducing any explicit time dependence in the Liouvillians. 
	
\begin{figure}
		\centering
		\includegraphics[width=\linewidth]{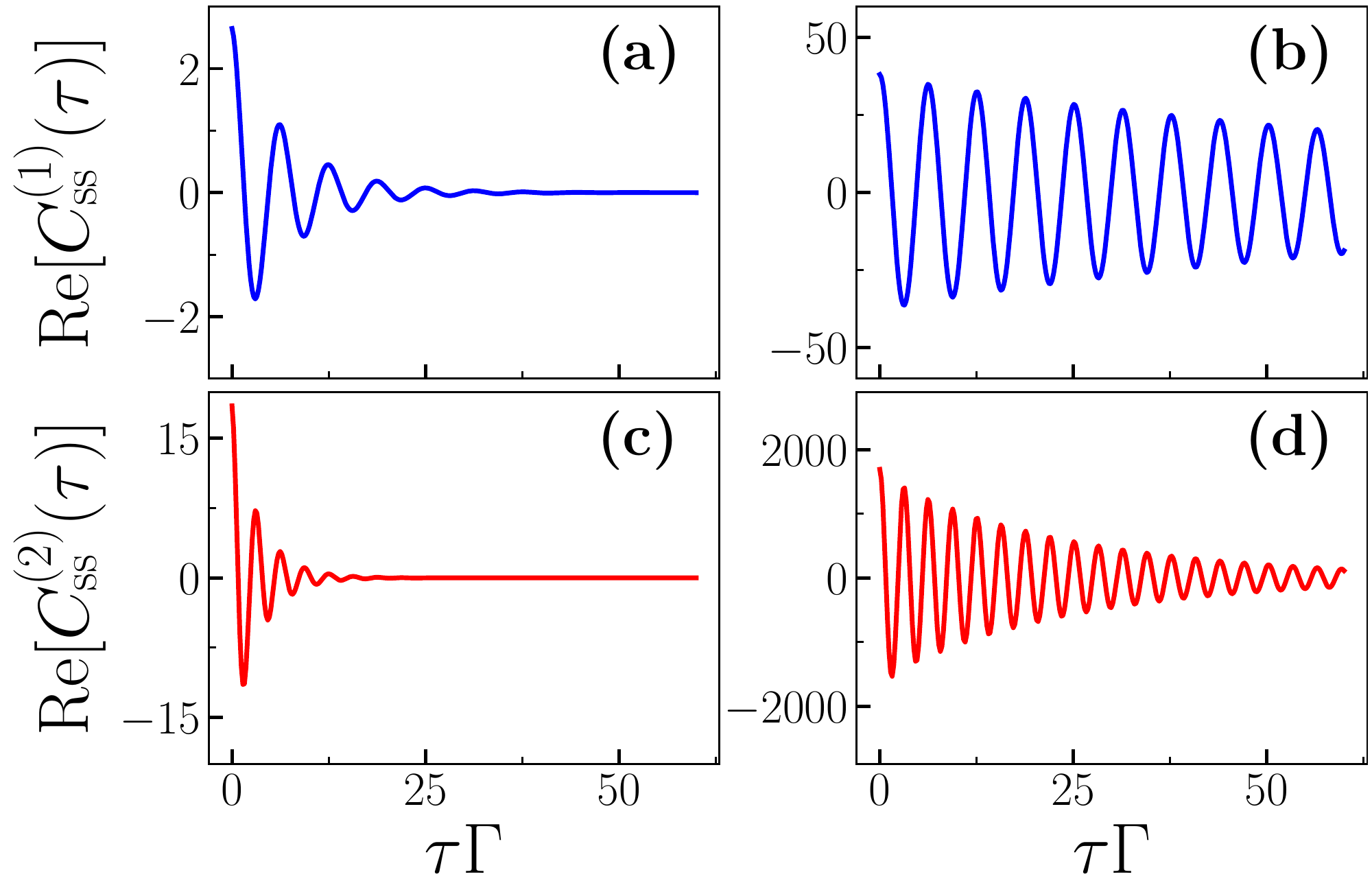}
		\caption{Real part of the two-time correlation functions $C^{(1)}_{\rm ss}(\tau)$ [(a) and (b)] and $C^{(2)}_{\rm ss}(\tau)$ [(c) and (d)] vs delay time $\tau$ in the steady state for: (a, c)  $\xi/\Gamma=0.75$ (before the DPT); (b, d) $\xi/\Gamma=1.25$ (after the DPT). Parameters: $\omega_c/\Gamma=1$, $\eta/\Gamma=0.1$, and $N=30$.}
		\label{fig:TTCF}
	\end{figure}

	
	The onset of DPTs can be visualized using quantum trajectory approaches \cite{BartoloEPJST17,JinPRB18,VicentiniPRA18} or by studying the dynamics of a properly initialized system \cite{LandaPRL20, LandaArXiv20}. Finally, we stress that both DPT and BTC can be observed by two-time correlation measures which have already been employed in the study of DPTs \cite{FinkNatPhys18}.
	In Figs.~\ref{fig:TTCF}(a) and \ref{fig:TTCF}(b) we plot two-time correlation function in the L-frame steady state $C^{(1)}_{\rm ss}(\tau)=\expec{\hat{a}^\dagger(0) \hat{a}(\tau)}_{\rm ss}$. The long-lasting coherence time associated to an oscillatory behaviour is the proof of the emergence of a BTC. 
	The abrupt increase in the amplitude of the oscillations demonstrates the emergence of the DPT.
	While $C^{(1)}_{\rm ss}(\tau)$ exhibits oscillations at frequency $\omega_c$, higher-order correlation functions can unveil those at multiple frequencies $k\omega_c$, $k\in\mathbb{N}$.
	For example, in Figs.~\ref{fig:TTCF}(c) and \ref{fig:TTCF}(d) we plot the two-time correlation function $C^{(2)}_{\rm ss}(\tau)=\expec{\hat{a}^{\dagger 2}(0)\hat{a}^2(\tau)}_{\rm ss}$. We stress that $C^{(2)}_{\rm ss}\neq g_{\rm ss}^{(2)}$ due to the different ordering of the boson operators. This specific ordering in $C^{(2)}_{\rm ss}$ allows one to observe the system oscillations frequency doubling those in  Figs.~\ref{fig:TTCF}(a) and \ref{fig:TTCF}(b).

	\begin{figure}
		\centering
		\includegraphics[width=0.95 \linewidth]{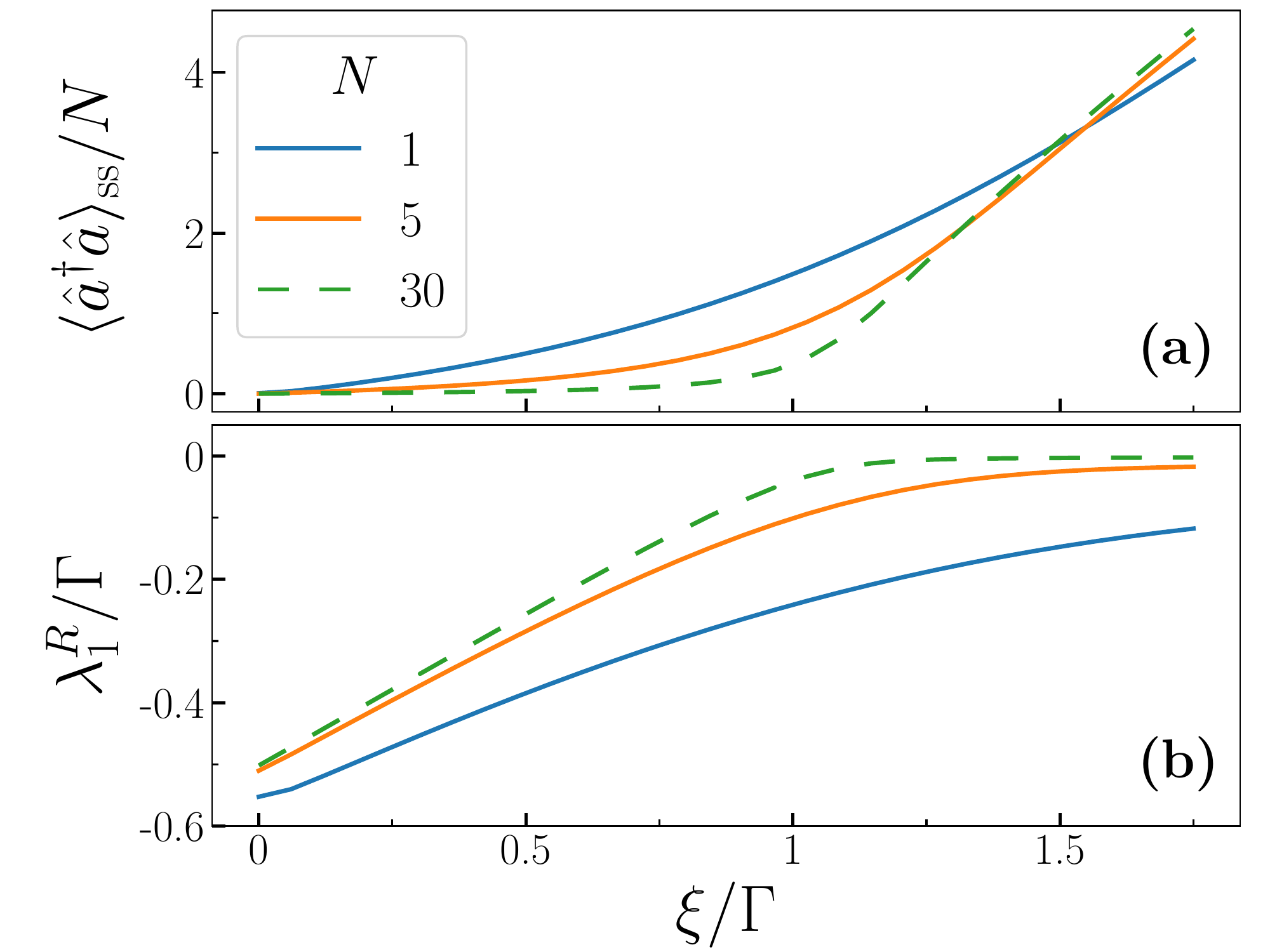}
		\caption{Numerical study of the onset of the DPT in the R-frame for the Scully-Lamb model.
			(a): Rescaled number of photons $\expec{\hat{a}^\dagger \hat{a}}_{\rm ss}/N$ as a function of the incoherent drive strength $\xi/\Gamma$.
			(b): Liouvillian gap $\lambda_1^R$ (in units of the damping rate $\Gamma$) as a function of the incoherent drive $\xi/\Gamma$.
			Parameters: $\eta/\Gamma=0.1$, $\beta/\Gamma=0.005$.
		}
		\label{fig:phase_transition_SL}
	\end{figure}

	\begin{figure}
		\centering
		\includegraphics[width=\linewidth]{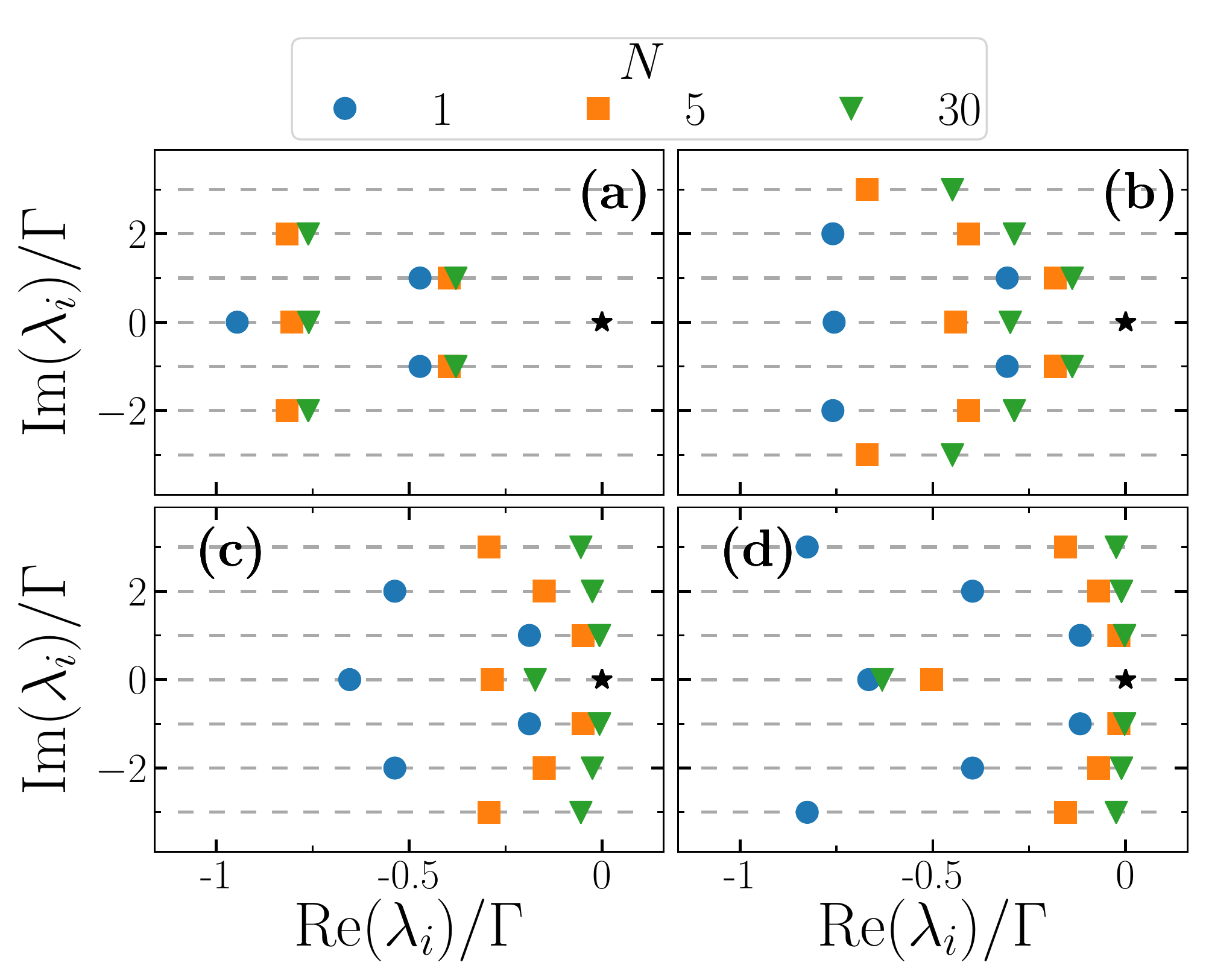}
		\caption{Real and imaginary parts of the Liouvillian eigenvalues for the Scully-Lamb model in the L-frame for $\xi/\Gamma$ equal to: (a) $0.25$, (b) $0.75$, (c) $1$, (d) $1.5$. Different markers represent different parameter $N$, while the black stars represent the steady-state eigenvalue $\lambda_0=0$. The gray horizontal dashed lines represent $\operatorname{Im}(\lambda_i)=k \omega_c$, where $k$ is an integer. The parameters used are: $\omega_c/\Gamma=1$, $\eta/\Gamma=0.1$, and $\beta/\Gamma=0.005$.}
		\label{fig:time_crystal_SL}
	\end{figure}

	\section{The Scully-Lamb laser model}
\label{Sec:Model2}
	The above results regarding the DPT-BTC correspondence can be extended to other models characterized by a $U(1)$ symmetry and which obey \eqref{Eq:Fundamental_correspondence}. 
	Here, we briefly show the emergence of the same phenomena in the celebrated Scully-Lamb laser model \cite{YamamotoBook,TakemuraArXiv19,ArkhipovPRA20} , whose Hamiltonian reads
	\begin{equation}
	\hat{H}=\omega_c \hat{a}^\dagger \hat{a}\,,
	\end{equation} 
	while the operators $\hat{L}_j$ are
	\begin{equation}\label{L123_SL}  
	\begin{split}
	&\hat L_1 = \ap \!\left(\!\sqrt{\xi}-\sqrt{\beta}\,\am\ap\!\right), \; \hat L_2 = \sqrt{\eta}\,\am\ap, \; \hat L_3 = \sqrt{\Gamma}\,\am,  
	\end{split}
	\end{equation}
	where $\hat{L}_1$ describes a \emph{nonlinear incoherent} drive (gain), $\hat{L}_2$ captures the nonlinear field decoherence, and $\hat L_3$ represents the single-particle loss.
	The parameter $\xi$ represents the medium gain (incoherent drive) strength, $\beta$ is the gain saturation (incoherent interaction) rate, $\eta$ is the decoherence rate, and $\Gamma$ is the inverse of the photon lifetime. 
	
	Such a somewhat simplified model can be obtained from the full Scully-Lamb laser master equation \cite{YamamotoBook} in the fourth-order field approximation, where $\xi= A$, $\beta=B^2/(4A)$ and $\eta=3B/4$, with $A$ and $B$ being the laser gain and gain saturation parameters, respectively (see \cite{Arkhipov2019,ArkhipovPRA20}). 
	Here, however, we consider $\beta$ and $\eta$ as independent parameters.
	To be physically meaningful, we must consider the weak gain-saturation regime, for which 
	$\sqrt{\beta/\xi}\ll 1$. Away from this limit, the system may become unstable.

	Similarly to the model in Sec.~\ref{Sec:Model}, we can introduce the R-frame Liouvillian $\LL^R$. Moreover, Eq.~\eqref{Eq:Fundamental_correspondence} remains valid also for this model.
	
	To investigate criticality, we introduce the scaling as a function of the parameter $N$ as $\{\xi, \Gamma, \eta, \beta\} \to \{\xi , \Gamma,  \eta/N, \beta/N^{2}\}$.
	In Fig.~\ref{fig:phase_transition_SL}, we show the onset of the transition for the R-frame Liouvillian $\LL^R$ as a function of $N$.
	Moreover, Fig.~\ref{fig:time_crystal_SL} demonstrate the emergence of a BTC, proving again the DPT-BTC correspondence.
	
	These results generalize and support the findings of Ref.~\cite{TakemuraArXiv19}, by providing a genuine interpretation of the well-known lasing transition~\cite{DeGiorgio1970} in a more general framework, by explicitly taking into account the symmetries of the model and the
	spectral properties of the Liouvillian.

	\section{Conclusions}
	\label{Sec:Conclusion}
	In this article, we have studied the emergence of a DPT and a BTC in a nonlinear optical model, where the nonlinearity comes from a dissipative term.
	To show these effects, we have analyzed the corresponding Lindblad-type
	Liouvillian superoperator and its spectrum for the studied system.
	
	We have shown that, in the thermodynamic limit,
	a second-order DPT in the
	R-frame corresponds to a BTC in the
	L-frame. 
	The two phenomena are the same
	in terms of the Liouvillian spectrum (or its gap) but
	just presented in different representations (frames).
	We find this prediction as the most important result of
	the manuscript. Moreover, \eqref{Eq:Fundamental_correspondence} is an indirect proof of the existence of time crystallization in open quantum systems, since it connects BTCs' existence to that of second-order DPTs.
	
	Beyond the interest in this correspondence, the novelty of the DPT discussed here is the \emph{incoherent} nature of all the processes involved.
	Indeed, the system studied in Refs.~\cite{CasteelsPRA17,BartoloPRA16} was a single Kerr resonator coherently driven, while in Ref.~\cite{BiellaPRA17} the incoherent (thermal-like) injection of photons was counterbalanced by a coherent two-body interaction process. Meaning that, in all these previous models, the system nonlinearity has been induced solely by a {\it coherent} interaction term, which is not the case for our model.
	This system can be realized with present technologies, such as in a incoherently driven engineered superconducting resonator \cite{LeghtasScience15,LescanneNatPhys2020}.
	
	Differently from Ref.~\cite{BucaNat19}, the example shown here does not admit mixed coherences through which a dark Hamiltonian can produce oscillations.
	It is the incoherent injection of photons leading to multiple symmetry-breaking steady states which permit the presence of everlasting oscillations.
	Moreover, contrary to Ref.~\cite{IeminiPRL18}, the presence of strong dissipative processes does not prevent crystallization.
	Since the emergence of a BTC is intertwined to the emergence of a symmetry-breaking DPT, the BTC discussed here is a critical phenomenon.
	The extension of these results to other types of symmetries is one of the perspectives of this work.
	
	The inclusion of nonlinear $U(1)$ Hamiltonian processes is straightforward, and allows to manipulate the time crystal oscillation periods. For example, by including Kerr-type nonlinearities the Liouvillian eigenvalues become not-equidistant and the resulting time crystals can have incommensurable oscillations. This will be the topic of future works.
	
	Finally, this article prompts the question of the existence of a similar transformation in different symmetric systems, allowing to reinterpret the appearance of a BTC as the physics of a DPT, but just in a generalized ``rotating frame''. Accordingly, emergent symmetries may be a key concept in understanding the physics of BTCs.

\begin{acknowledgments}
		The authors acknowledge the critical reading of Marcello Dalmonte, Simone Felicetti, and Vincenzo Macr\'i.
		The authors are grateful to the RIKEN Advanced Center for Computing and Communication (ACCC) for the allocation of computational resources of the RIKEN supercomputer system (HOKUSAI BigWaterfall).  
		I.A. thanks the Grant Agency of the Czech
		Republic (Project No.~18-08874S), and Project no.
		CZ.02.1.01\/0.0\/0.0\/16\_019\/0000754 of the Ministry of
		Education, Youth and Sports of the Czech Republic. A.M. is
		supported by the Polish National Science Centre (NCN) under the
		Maestro Grant No. DEC-2019/34/A/ST2/00081. F.N. is supported in
		part by: NTT Research, Army Research Office (ARO) (Grant No.
		W911NF-18-1-0358), Japan Science and Technology Agency (JST) (via
		Q-LEAP and the CREST Grant No. JPMJCR1676), Japan Society for the
		Promotion of Science (JSPS) (via the KAKENHI Grant No. JP20H00134,
		and the JSPS-RFBR Grant No. JPJSBP120194828), and the Grant No.
		FQXi-IAF19-06 from the Foundational Questions Institute Fund
		(FQXi), a donor advised fund of the Silicon Valley Community
		Foundation.
\end{acknowledgments}

\appendix

\section{Properties of Liouvillian symmetries}
\label{App:Properties}

The Lindblad master equation, given in  \eqref{Eq:Lindblad1} is invariant under any transformation $\hat{a}\to \hat{a}\exp({i \phi})$ for an arbitrary real number $\phi$.
Thus, the model exhibits the $U(1)$ symmetry and the superoperator $\mathcal{U}$, defined as
\begin{equation}\label{Eqapp:symmetry}
\mathcal{U} \rhot = \exp\left({-i \phi \hat{a}^\dagger \hat{a}}\right) \rhot \exp\left({i \phi \hat{a}^\dagger \hat{a}}\right),
\end{equation}
commutes with the Liouvillian: $[\LL, \mathcal{U}]=0$.

While Liouvillian symmetries  are not all associated to conserved quantities \cite{BucaNPJ2012,AlbertPRA14,BaumgartnerJPA08}, they still constrain the system dynamics.
Indeed, all the eigenmatrices of $\mathcal{L}$ must be eigenmatrices of $\mathcal{U}$, that is, 
\begin{equation}\label{Eq:symmetry_on_eigenmatrices}
\mathcal{U} \eig{i} = u_i \eig{i},    
\end{equation} 
where $u_i$ is an eigenvalue of $\mathcal{U}$  \cite{BaumgartnerJPA08,MingantPRA18_Spectral,AlbertPRA14,AlbertPRX16}.

The span of all the eigenmatrices $\eig{i}$ with the same $u_i$, such that 
\begin{equation}
\LL \eig{i}=u_i \eig{i},
\end{equation} 
defines a \emph{symmetry sector}. 
Each symmetry sector represents a part of the Liouvillian space which is not connected to its
other parts (sectors) by the Liouvillian dynamics.

To better grasp the meaning of this symmetry, let us express the eigenmatrix $\eig{i}$ in the number (Fock) basis as
\begin{equation}\label{rho_i}
\hat\rho_i=\sum_{m,n} c_{m,n}\ket{m}\bra{n}.
\end{equation}
By combining Eqs.~(\ref{Eqapp:symmetry}), (\ref{Eq:symmetry_on_eigenmatrices}),~and~(\ref{rho_i}), one obtains 
\begin{equation}
\begin{split}
\mathcal{U}\eig{i} &= \sum_{m,n} c_{m,n} e^{-i \phi \hat{a}^\dagger \hat{a}}  \ket{m}\bra{n} e^{i \phi \hat{a}^\dagger \hat{a}} \\
& =\sum_{m,n} c_{m,n} e^{-i \phi (m-n)}\ket{m}\bra{n}= u_i \eig{i}.
\end{split}
\end{equation}
We conclude that $\exp\left[{-i \phi (m-n)}\right]$ must be a constant and, therefore, any eigenmatrix $\hat\rho_i$ in Eq.~(\ref{rho_i}) must obey
\begin{equation}\label{Eq:condition_symmetry}
\eig{i}= \sum_{m} c_{m} \ket{m}\bra{m-k} \, ,
\end{equation}
for some constant integer $k\in \mathbb Z$.
In other words, $\eig{i}$ must be an operator containing elements only on one diagonal, and different symmetry sectors occupy different upper and lower diagonals.

\section{Proof of Eq. (7)}
\label{App:Proof}

We can eliminate the Hamiltonian dependence of the Liouvillian by choosing the frame which rotates at the cavity frequency (the R-frame), i.e., the one rotating at the cavity frequency $\omega_c$.
The density matrix in the R-frame is
\begin{equation}
\hat{\rho}^R(t) = \exp\left({i \omega_c t \hat{a}^\dagger \hat{a}}\right) \rhot  \exp\left({-i \omega_c t \hat{a}^\dagger \hat{a}}\right).
\end{equation}

We notice that, since all the dissipators are $U(1)$-symmetric and the rotation is equivalent to applying the symmetry superoperator $\mathcal{U}$ with $\phi=\omega_c t$, the dissipators are unchanged by the transformation. Hence, the Liouvillian in the R-frame is
\begin{equation}
\label{Eq:Liouvillian_without_Hamiltonian}
\partial_t \hat{\rho}^R(t)= \LL^R \hat{\rho}^R(t) = \left( \DD[\hat{L}_1] + \DD[\hat{L}_2] +\DD[\hat{L}_3] \right) \hat{\rho}^R(t).
\end{equation}

Having introduced the R-frame Liouvillian $\LL^R$, we can introduce its eigenvalues $\lambda_i^{R}$ and eigenvectors $\eig{i}^{R}$ defined by
\begin{equation}
\LL^R \eig{i}^{R} =\lambda^R \eig{i}^{R}.
\end{equation}
Notice that $\LL^R$ has exactly the same symmetries of the original problem, since $[\mathcal{U}, \LL^{R}]=0$. 
Thus, the condition in \eqref{Eq:condition_symmetry} remains valid also for $\eig{i}^R$, i.e., 
\begin{equation}\label{Eq:condition_symmetry_2}
\eig{i}^R= \sum_{p} c_{p} \ket{p}\bra{p-k}.
\end{equation}
Also note that, due to the super- and sub-diagonal form of $\hat\rho_i^R$ in Eq.~(\ref{Eq:condition_symmetry_2}), the  eigenmatrices $\hat\rho_i^R$ are, in general, non-Hermitian. This means that the corresponding eigenvalues $\lambda_i^R$ are, in general, complex. 

For a generic model and a generic change of reference, there is no trivial correspondence 
between $\lambda_i^{R}$ and $\lambda_i$, as well as $\eig{i}$ and $\eig{i}^{R}$.
However, for the model under consideration we have
\begin{equation}\begin{split}
\LL \eig{i}^{R} &= \LL^R \eig{i}^{R} - i \omega_c \left[\hat{a}^\dagger \hat{a}, \eig{i}^{R}\right] \\ &= \lambda_{i}^{R} \eig{i}^{R} - i \omega_c \sum_{p} c_{p}  \left[\hat{a}^\dagger \hat{a}, \ket{p}\bra{p-k}\right] \\ &= \lambda_{i}^{R} \eig{i}^{R} - i \omega_c k \sum_{p} c_{p}  \ket{p}\bra{p-k} \\ &= (\lambda_{i}^{R} - i \omega_c k) \eig{i}^{R},
\end{split}
\end{equation}
where the latter follows from \eqref{Eq:condition_symmetry_2}, with $k$ an integer number.
Thus, we have proved  the following fundamental equalities:
\begin{equation}\label{Eq:Fundamental_correspondence_app}
\eig{i}= \eig{i}^{R}\, , \quad \lambda_i = \lambda_i^{R} - i \omega_c k. 
\end{equation}
This is exactly  \eqref{Eq:Fundamental_correspondence}.

\section{Thermodynamic limit of a single bosonic cavity}
\label{App:Thermodynamic_limit}

To grasp the correct scaling towards the thermodynamic limit, we can consider the semiclassical equation of motion and search  for that scaling of the parameters under which  $n\to N n$, as detailed in Ref.~\cite{CasteelsPRA17}. 
Any transformations $\{\Gamma, \xi, \eta\} \to \{N^{\mu} \Gamma, N^{\mu} \xi , \eta/N^{1-\mu}\}$, for arbitrary $\mu$, respect the photon number scaling $n\to N n$, but only $\mu=2$ is physically meaningful. Indeed, for any $\mu\neq 0$, the convergence towards the steady state becomes faster (or slower) for $\xi=0$. Equivalently, the natural timescale of the problem, which is given by the photon-dissipation rate $\Gamma$, should not be modified by increasing the system size.

A different argument can be provided by considering a lattice of resonators:
\begin{equation}
\hat{H} =\omega_c \sum_{i=1}^{N} \hat{a}^\dagger_i \hat{a}_i+  J \sum_{\expec{i, j}}  \hat{a}^\dagger_{i} \hat{a}_j,
\end{equation}
where $\expec{i, j}$ indicates that the sum runs over the nearest neighbours.
Note that this Hamiltonian includes the $J$ term and, thus, it is more
general than that considered in Sec.~\ref{Sec:Model}.
The corresponding Liouvillian reads
\begin{equation}
\begin{split}
\LL \rhot = -i &\left[ \hat{H}, \rhot \right] +  \DD\left[\sum_{i=1}^{N} \sqrt{\frac{\xi}{N}}\hat{a}^\dagger_{i} \right] \\
& + \sum_{i=1}^{N} \left( \DD\left[\sqrt{\Gamma}\hat{a}_{i} \right]+  \DD\left[\sqrt{\eta}\hat{a}^2_{i} \right] \right) \rhot.
\end{split}
\end{equation}
Notice the fundamental difference between $\DD\left[\sum_i \hat{L}_i \right]$ and $\sum_i  \DD\left[\hat{L}_i \right]$.
Indeed, in this model we are assuming that every cavity is identical, but while photonic emission is not correlated, the incoherent drive is only in the uniform mode of the cavity.

In the momentum space the Hamiltonian and dissipators are
\begin{equation}
\begin{split}
\hat{H} &= \sum_k \left[\omega_c- 2J \cos(k)\right]\hat{a}^\dagger_{k} \hat{a}_k\, , \\ 
\sum_i \DD\left[\sqrt{\Gamma}\hat{a}_{i} \right] &= \sum_k\DD\left[\sqrt{\Gamma}\hat{a}_{k} \right],\\
\DD\left[\sum_{i=1}^{N} \sqrt{\frac{\xi}{N}}\hat{a}^\dagger_{i} \right] &= \sum_k\DD\left[\sqrt{\xi}\hat{a}_{k}^\dagger \right], \\
\sum_i \DD\left[\sqrt{\eta}\hat{a}_{i}^2 \right] &= \frac{\xi}{N}\sum_{k,k', q}\DD\left[\hat{a}_{k+q}\hat{a}_{k'-q};\,\hat{a}_{k}^\dagger\hat{a}_{k'}^\dagger \right],
\end{split}
\end{equation}
where $\DD\left[\hat{A}, \hat{B}\right]= \hat{A} \bigcdot \hat{B} -\frac{1}{2}(\hat{A}  \hat{B} \bigcdot + \bigcdot \hat{A} \hat{B})$ is the generalized Lindblad dissipator, and $\bigcdot$ is the placeholder for the density matrix [e.g., $(\hat{A} \bigcdot \hat{B}   ) \rhot= \hat{A} \rhot \hat{B}   $].
For a large $N$, the effect of $\eta$ is vanishingly small. Nevertheless, we cannot neglect it, since it is the dominant term in the phase with a large number of photons.
However, we may effectively decouple the mode for $k=0$ (the only driven one) from those of the other modes.
The resulting model is that provided in \eqref{Lindblad1}, whose scaling with the parameter $N$ is $\{\Gamma, \xi, \eta\} \to \{\Gamma,  \xi , \eta/N\}$.

$\ $ \\
$\ $ \\
$\ $ \\
$\ $ \\


\begin{thebibliography}{81}%
	\makeatletter
	\providecommand \@ifxundefined [1]{%
		\@ifx{#1\undefined}
	}%
	\providecommand \@ifnum [1]{%
		\ifnum #1\expandafter \@firstoftwo
		\else \expandafter \@secondoftwo
		\fi
	}%
	\providecommand \@ifx [1]{%
		\ifx #1\expandafter \@firstoftwo
		\else \expandafter \@secondoftwo
		\fi
	}%
	\providecommand \natexlab [1]{#1}%
	\providecommand \enquote  [1]{``#1''}%
	\providecommand \bibnamefont  [1]{#1}%
	\providecommand \bibfnamefont [1]{#1}%
	\providecommand \citenamefont [1]{#1}%
	\providecommand \href@noop [0]{\@secondoftwo}%
	\providecommand \href [0]{\begingroup \@sanitize@url \@href}%
	\providecommand \@href[1]{\@@startlink{#1}\@@href}%
	\providecommand \@@href[1]{\endgroup#1\@@endlink}%
	\providecommand \@sanitize@url [0]{\catcode `\\12\catcode `\$12\catcode
		`\&12\catcode `\#12\catcode `\^12\catcode `\_12\catcode `\%12\relax}%
	\providecommand \@@startlink[1]{}%
	\providecommand \@@endlink[0]{}%
	\providecommand \url  [0]{\begingroup\@sanitize@url \@url }%
	\providecommand \@url [1]{\endgroup\@href {#1}{\urlprefix }}%
	\providecommand \urlprefix  [0]{URL }%
	\providecommand \Eprint [0]{\href }%
	\providecommand \doibase [0]{http://dx.doi.org/}%
	\providecommand \selectlanguage [0]{\@gobble}%
	\providecommand \bibinfo  [0]{\@secondoftwo}%
	\providecommand \bibfield  [0]{\@secondoftwo}%
	\providecommand \translation [1]{[#1]}%
	\providecommand \BibitemOpen [0]{}%
	\providecommand \bibitemStop [0]{}%
	\providecommand \bibitemNoStop [0]{.\EOS\space}%
	\providecommand \EOS [0]{\spacefactor3000\relax}%
	\providecommand \BibitemShut  [1]{\csname bibitem#1\endcsname}%
	\let\auto@bib@innerbib\@empty
	\bibitem [{\citenamefont {Kane}\ and\ \citenamefont {Mele}(2005)}]{Kane2005}%
	\BibitemOpen
	\bibinfo {author} {C.~L. Kane}\ and\ \bibinfo {author} {E.~J. Mele},\ \emph
	{\bibinfo {title} {${Z}_{2}$ Topological Order and the Quantum Spin {H}all
			Effect}},\ \href {\doibase 10.1103/PhysRevLett.95.146802} {\bibfield
		{journal} {\bibinfo  {journal} {Phys. Rev. Lett.}\ }\textbf {\bibinfo
			{volume} {95}},\ \bibinfo {pages} {146802} (\bibinfo {year}
		{2005})}\BibitemShut {NoStop}%
	\bibitem [{\citenamefont {Maldacena}(1999)}]{Maldacena1999}%
	\BibitemOpen
	\bibinfo {author} {J.~Maldacena},\ \emph {\bibinfo {title} {The Large-{N}
			Limit of Superconformal Field Theories and Supergravity}},\ \href {\doibase
		10.1023/a:1026654312961} {\bibfield  {journal} {\bibinfo  {journal} {Int. J.
				Theor. Phys.}\ }\textbf {\bibinfo {volume} {38}},\ \bibinfo {pages} {1113}
		(\bibinfo {year} {1999})}\BibitemShut {NoStop}%
	\bibitem [{\citenamefont {Carmichael}(2015)}]{CarmichaelPRX15}%
	\BibitemOpen
	\bibinfo {author} {H.~J. Carmichael},\ \emph {\bibinfo {title} {Breakdown of
			Photon Blockade: A Dissipative Quantum Phase Transition in Zero
			Dimensions}},\ \href {https://link.aps.org/doi/10.1103/PhysRevX.5.031028}
	{\bibfield  {journal} {\bibinfo  {journal} {Phys. Rev. X}\ }\textbf {\bibinfo
			{volume} {5}},\ \bibinfo {pages} {031028} (\bibinfo {year}
		{2015})}\BibitemShut {NoStop}%
	\bibitem [{\citenamefont {Fink}\ \emph {et~al.}(2018)\citenamefont {Fink},
		\citenamefont {Schade}, \citenamefont {H{\"o}fling}, \citenamefont
		{Schneider},\ and\ \citenamefont {Imamoglu}}]{FinkNatPhys18}%
	\BibitemOpen
	\bibinfo {author} {T.~Fink}, \bibinfo {author} {A.~Schade}, \bibinfo {author}
	{S.~H{\"o}fling}, \bibinfo {author} {C.~Schneider},\ and\ \bibinfo {author}
	{A.~Imamoglu},\ \emph {\bibinfo {title} {Signatures of a dissipative phase
			transition in photon correlation measurements}},\ \href
	{https://doi.org/10.1038/s41567-017-0020-9} {\bibfield  {journal} {\bibinfo
			{journal} {Nature Physics}\ }\textbf {\bibinfo {volume} {14}},\ \bibinfo
		{pages} {365} (\bibinfo {year} {2018})}\BibitemShut {NoStop}%
	\bibitem [{\citenamefont {Minganti}\ \emph {et~al.}(2018)\citenamefont
		{Minganti}, \citenamefont {Biella}, \citenamefont {Bartolo},\ and\
		\citenamefont {Ciuti}}]{MingantPRA18_Spectral}%
	\BibitemOpen
	\bibinfo {author} {F.~Minganti}, \bibinfo {author} {A.~Biella}, \bibinfo
	{author} {N.~Bartolo},\ and\ \bibinfo {author} {C.~Ciuti},\ \emph {\bibinfo
		{title} {Spectral theory of Liouvillians for dissipative phase
			transitions}},\ \href {https://link.aps.org/doi/10.1103/PhysRevA.98.042118}
	{\bibfield  {journal} {\bibinfo  {journal} {Phys. Rev. A}\ }\textbf {\bibinfo
			{volume} {98}},\ \bibinfo {pages} {042118} (\bibinfo {year}
		{2018})}\BibitemShut {NoStop}%
	\bibitem [{Note1()}]{Note1}%
	\BibitemOpen
	\bibinfo {note} {The name boundary time crystals is used in Refs.~\cite
		{IeminiPRL18, LLedoNJP20} as a synonym of dissipative time crystal. Indeed, a
		dissipative system can be seen as the boundary of the universe (where the
		universe is the sum of the system and its environment). Tracing out all the
		degrees of freedom of the environment, the boundary time crystals
		emerge.}\BibitemShut {Stop}%
	\bibitem [{\citenamefont {Iemini}\ \emph {et~al.}(2018)\citenamefont {Iemini},
		\citenamefont {Russomanno}, \citenamefont {Keeling}, \citenamefont
		{Schir\`o}, \citenamefont {Dalmonte},\ and\ \citenamefont
		{Fazio}}]{IeminiPRL18}%
	\BibitemOpen
	\bibinfo {author} {F.~Iemini}, \bibinfo {author} {A.~Russomanno}, \bibinfo
	{author} {J.~Keeling}, \bibinfo {author} {M.~Schir\`o}, \bibinfo {author}
	{M.~Dalmonte},\ and\ \bibinfo {author} {R.~Fazio},\ \emph {\bibinfo {title}
		{Boundary Time Crystals}},\ \href {\doibase 10.1103/PhysRevLett.121.035301}
	{\bibfield  {journal} {\bibinfo  {journal} {Phys. Rev. Lett.}\ }\textbf
		{\bibinfo {volume} {121}},\ \bibinfo {pages} {035301} (\bibinfo {year}
		{2018})}\BibitemShut {NoStop}%
	\bibitem [{\citenamefont {Lled\'o}\ and\ \citenamefont
		{Szyma\'{n}ska}(2020)}]{LLedoNJP20}%
	\BibitemOpen
	\bibinfo {author} {C.~Lled\'o}\ and\ \bibinfo {author} {M.~H.
		Szyma\'{n}ska},\ \emph {\bibinfo {title} {Dissipative time crystal with or
			without $Z_2$ symmetry breaking}},\ \href
	{http://iopscience.iop.org/10.1088/1367-2630/ab9ae3} {\bibfield  {journal}
		{\bibinfo  {journal} {New Journal of Physics}\ } (\bibinfo {year}
		{2020})}\BibitemShut {NoStop}%
	\bibitem [{\citenamefont {Carusotto}\ and\ \citenamefont
		{Ciuti}()}]{Carusotto_RMP_2013_quantum_fluids_light}%
	\BibitemOpen
	\bibinfo {author} {I.~Carusotto}\ and\ \bibinfo {author} {C.~Ciuti},\ \emph
	{\bibinfo {title} {Quantum fluids of light}},\ \href
	{https://link.aps.org/doi/10.1103/RevModPhys.85.299} {\bibfield  {journal}
		{\bibinfo  {journal} {Rev. Mod. Phys.}\ }\textbf {\bibinfo {volume} {85}},\
		\bibinfo {pages} {299}}\BibitemShut {NoStop}%
	\bibitem [{\citenamefont {Kockum}\ \emph {et~al.}(2019)\citenamefont {Kockum},
		\citenamefont {Miranowicz}, \citenamefont {De~Liberato}, \citenamefont
		{Savasta},\ and\ \citenamefont {Nori}}]{kockum2019}%
	\BibitemOpen
	\bibinfo {author} {A.~F. Kockum}, \bibinfo {author} {A.~Miranowicz}, \bibinfo
	{author} {S.~De~Liberato}, \bibinfo {author} {S.~Savasta},\ and\ \bibinfo
	{author} {F.~Nori},\ \emph {\bibinfo {title} {Ultrastrong coupling between
			light and matter}},\ \href {\doibase
		https://doi.org/10.1038/s42254-018-0006-2} {\bibfield  {journal} {\bibinfo
			{journal} {Nat. Rev. Phys.}\ }\textbf {\bibinfo {volume} {1}},\ \bibinfo
		{pages} {19} (\bibinfo {year} {2019})}\BibitemShut {NoStop}%
	\bibitem [{\citenamefont {Delteil}\ \emph {et~al.}(2019)\citenamefont
		{Delteil}, \citenamefont {Fink}, \citenamefont {Schade}, \citenamefont
		{H{\"o}fling}, \citenamefont {Schneider},\ and\ \citenamefont
		{Imamoglu}}]{DelteilNatMat19}%
	\BibitemOpen
	\bibinfo {author} {A.~Delteil}, \bibinfo {author} {T.~Fink}, \bibinfo
	{author} {A.~Schade}, \bibinfo {author} {S.~H{\"o}fling}, \bibinfo {author}
	{C.~Schneider},\ and\ \bibinfo {author} {A.~Imamoglu},\ \emph {\bibinfo
		{title} {Towards polariton blockade of confined exciton-polaritons}},\ \href
	{\doibase 10.1038/s41563-019-0282-y} {\bibfield  {journal} {\bibinfo
			{journal} {Nature Materials}\ }\textbf {\bibinfo {volume} {18}},\ \bibinfo
		{pages} {219} (\bibinfo {year} {2019})}\BibitemShut {NoStop}%
	\bibitem [{\citenamefont {You}\ and\ \citenamefont {Nori}(2011)}]{YouNat11}%
	\BibitemOpen
	\bibinfo {author} {J.~Q. You}\ and\ \bibinfo {author} {F.~Nori},\ \emph
	{\bibinfo {title} {Atomic physics and quantum optics using superconducting
			circuits}},\ \href {http://dx.doi.org/10.1038/nature10122} {\bibfield
		{journal} {\bibinfo  {journal} {Nature (London)}\ }\textbf {\bibinfo {volume}
			{474}},\ \bibinfo {pages} {589} (\bibinfo {year} {2011})}\BibitemShut
	{NoStop}%
	\bibitem [{\citenamefont {Gu}\ \emph {et~al.}(2017)\citenamefont {Gu},
		\citenamefont {Kockum}, \citenamefont {Miranowicz}, \citenamefont {Liu},\
		and\ \citenamefont {Nori}}]{Gu2017}%
	\BibitemOpen
	\bibinfo {author} {X.~Gu}, \bibinfo {author} {A.~F. Kockum}, \bibinfo
	{author} {A.~Miranowicz}, \bibinfo {author} {Y.~X. Liu},\ and\ \bibinfo
	{author} {F.~Nori},\ \emph {\bibinfo {title} {Microwave photonics with
			superconducting quantum circuits}},\ \href
	{https://doi.org/10.1016/j.physrep.2017.10.002} {\bibfield  {journal}
		{\bibinfo  {journal} {Phys. Rep.}\ }\textbf {\bibinfo {volume} {718-719}},\
		\bibinfo {pages} {1} (\bibinfo {year} {2017})}\BibitemShut {NoStop}%
	\bibitem [{\citenamefont {Kjaergaard}\ \emph {et~al.}(2020)\citenamefont
		{Kjaergaard}, \citenamefont {Schwartz}, \citenamefont {Braum\"uller},
		\citenamefont {Krantz}, \citenamefont {Wang}, \citenamefont {Gustavsson},\
		and\ \citenamefont {Oliver}}]{Kjaergaard20}%
	\BibitemOpen
	\bibinfo {author} {M.~Kjaergaard}, \bibinfo {author} {M.~E. Schwartz},
	\bibinfo {author} {J.~Braum\"uller}, \bibinfo {author} {P.~Krantz}, \bibinfo
	{author} {J.~I.-J. Wang}, \bibinfo {author} {S.~Gustavsson},\ and\ \bibinfo
	{author} {W.~D. Oliver},\ \emph {\bibinfo {title} {Superconducting Qubits:
			Current State of Play}},\ \href {\doibase
		10.1146/annurev-conmatphys-031119-050605} {\bibfield  {journal} {\bibinfo
			{journal} {Annu. Rev. Condens. Matter Phys.}\ }\textbf {\bibinfo {volume}
			{11}},\ \bibinfo {pages} {369} (\bibinfo {year} {2020})}\BibitemShut
	{NoStop}%
	\bibitem [{\citenamefont {Breuer}\ and\ \citenamefont
		{Petruccione}(2007)}]{BreuerBookOpen}%
	\BibitemOpen
	\bibinfo {author} {H.~Breuer}\ and\ \bibinfo {author} {F.~Petruccione},\
	\href@noop {} {\emph {\bibinfo {title} {The Theory of Open Quantum
				Systems}}}\ (\bibinfo  {publisher} {Oxford University Press},\ \bibinfo
	{address} {Oxford},\ \bibinfo {year} {2007})\BibitemShut {NoStop}%
	\bibitem [{\citenamefont {Lidar}()}]{LidarLectureNotes}%
	\BibitemOpen
	\bibinfo {author} {D.~A. Lidar},\ \emph {\bibinfo {title} {Lecture Notes on
			the Theory of Open Quantum Systems}},\ \href
	{https://arxiv.org/abs/1902.00967} {\ }\Eprint
	{http://arxiv.org/abs/arXiv:1902.00967} {arXiv:1902.00967} \BibitemShut
	{NoStop}%
	\bibitem [{\citenamefont {DeGiorgio}\ and\ \citenamefont
		{Scully}(1970)}]{DeGiorgio1970}%
	\BibitemOpen
	\bibinfo {author} {V.~DeGiorgio}\ and\ \bibinfo {author} {M.~O. Scully},\
	\emph {\bibinfo {title} {Analogy between the Laser Threshold Region and a
			Second-Order Phase Transition}},\ \href {\doibase 10.1103/PhysRevA.2.1170}
	{\bibfield  {journal} {\bibinfo  {journal} {Phys. Rev. A}\ }\textbf {\bibinfo
			{volume} {2}},\ \bibinfo {pages} {1170} (\bibinfo {year} {1970})}\BibitemShut
	{NoStop}%
	\bibitem [{\citenamefont {Bonifacio}\ \emph {et~al.}(1978)\citenamefont
		{Bonifacio}, \citenamefont {Gronchi},\ and\ \citenamefont
		{Lugiato}}]{BonifacioPRA78}%
	\BibitemOpen
	\bibinfo {author} {R.~Bonifacio}, \bibinfo {author} {M.~Gronchi},\ and\
	\bibinfo {author} {L.~A. Lugiato},\ \emph {\bibinfo {title} {Photon
			statistics of a bistable absorber}},\ \href {\doibase
		10.1103/PhysRevA.18.2266} {\bibfield  {journal} {\bibinfo  {journal} {Phys.
				Rev. A}\ }\textbf {\bibinfo {volume} {18}},\ \bibinfo {pages} {2266}
		(\bibinfo {year} {1978})}\BibitemShut {NoStop}%
	\bibitem [{\citenamefont {Mollow}\ and\ \citenamefont
		{Glauber}(1967)}]{Mollow67}%
	\BibitemOpen
	\bibinfo {author} {B.~R. Mollow}\ and\ \bibinfo {author} {R.~J. Glauber},\
	\emph {\bibinfo {title} {Quantum Theory of Parametric Amplification. I}},\
	\href {\doibase 10.1103/PhysRev.160.1076} {\bibfield  {journal} {\bibinfo
			{journal} {Phys. Rev.}\ }\textbf {\bibinfo {volume} {160}},\ \bibinfo {pages}
		{1076} (\bibinfo {year} {1967})}\BibitemShut {NoStop}%
	\bibitem [{\citenamefont {Drummond}\ and\ \citenamefont
		{Carmichael}(1978)}]{Drummond1978}%
	\BibitemOpen
	\bibinfo {author} {P.~Drummond}\ and\ \bibinfo {author} {H.~Carmichael},\
	\emph {\bibinfo {title} {Volterra cycles and the cooperative fluorescence
			critical point}},\ \href {\doibase 10.1016/0030-4018(78)90198-0} {\bibfield
		{journal} {\bibinfo  {journal} {Opt. Commun.}\ }\textbf {\bibinfo {volume}
			{27}},\ \bibinfo {pages} {160} (\bibinfo {year} {1978})}\BibitemShut
	{NoStop}%
	\bibitem [{\citenamefont {Weimer}(2015)}]{WeimerPRL2015}%
	\BibitemOpen
	\bibinfo {author} {H.~Weimer},\ \emph {\bibinfo {title} {Variational
			Principle for Steady States of Dissipative Quantum Many-Body Systems}},\
	\href {https://link.aps.org/doi/10.1103/PhysRevLett.114.040402} {\bibfield
		{journal} {\bibinfo  {journal} {Phys. Rev. Lett.}\ }\textbf {\bibinfo
			{volume} {114}},\ \bibinfo {pages} {040402} (\bibinfo {year}
		{2015})}\BibitemShut {NoStop}%
	\bibitem [{\citenamefont {Mendoza-Arenas}\ \emph {et~al.}(2016)\citenamefont
		{Mendoza-Arenas}, \citenamefont {Clark}, \citenamefont {Felicetti},
		\citenamefont {Romero}, \citenamefont {Solano}, \citenamefont {Angelakis},\
		and\ \citenamefont {Jaksch}}]{MendozaPRA16}%
	\BibitemOpen
	\bibinfo {author} {J.~J. Mendoza-Arenas}, \bibinfo {author} {S.~R. Clark},
	\bibinfo {author} {S.~Felicetti}, \bibinfo {author} {G.~Romero}, \bibinfo
	{author} {E.~Solano}, \bibinfo {author} {D.~G. Angelakis},\ and\ \bibinfo
	{author} {D.~Jaksch},\ \emph {\bibinfo {title} {Beyond mean-field bistability
			in driven-dissipative lattices: Bunching-antibunching transition and quantum
			simulation}},\ \href {https://link.aps.org/doi/10.1103/PhysRevA.93.023821}
	{\bibfield  {journal} {\bibinfo  {journal} {Phys. Rev. A}\ }\textbf {\bibinfo
			{volume} {93}},\ \bibinfo {pages} {023821} (\bibinfo {year}
		{2016})}\BibitemShut {NoStop}%
	\bibitem [{\citenamefont {Casteels}\ \emph {et~al.}(2016)\citenamefont
		{Casteels}, \citenamefont {Storme}, \citenamefont {Le~Boit\'e},\ and\
		\citenamefont {Ciuti}}]{CasteelsPRA16}%
	\BibitemOpen
	\bibinfo {author} {W.~Casteels}, \bibinfo {author} {F.~Storme}, \bibinfo
	{author} {A.~Le~Boit\'e},\ and\ \bibinfo {author} {C.~Ciuti},\ \emph
	{\bibinfo {title} {Power laws in the dynamic hysteresis of quantum nonlinear
			photonic resonators}},\ \href
	{https://link.aps.org/doi/10.1103/PhysRevA.93.033824} {\bibfield  {journal}
		{\bibinfo  {journal} {Phys. Rev. A}\ }\textbf {\bibinfo {volume} {93}},\
		\bibinfo {pages} {033824} (\bibinfo {year} {2016})}\BibitemShut {NoStop}%
	\bibitem [{\citenamefont {Bartolo}\ \emph {et~al.}(2016)\citenamefont
		{Bartolo}, \citenamefont {Minganti}, \citenamefont {Casteels},\ and\
		\citenamefont {Ciuti}}]{BartoloPRA16}%
	\BibitemOpen
	\bibinfo {author} {N.~Bartolo}, \bibinfo {author} {F.~Minganti}, \bibinfo
	{author} {W.~Casteels},\ and\ \bibinfo {author} {C.~Ciuti},\ \emph {\bibinfo
		{title} {Exact steady state of a Kerr resonator with one- and two-photon
			driving and dissipation: Controllable Wigner-function multimodality and
			dissipative phase transitions}},\ \href
	{https://link.aps.org/doi/10.1103/PhysRevA.94.033841} {\bibfield  {journal}
		{\bibinfo  {journal} {Phys. Rev. A}\ }\textbf {\bibinfo {volume} {94}},\
		\bibinfo {pages} {033841} (\bibinfo {year} {2016})}\BibitemShut {NoStop}%
	\bibitem [{\citenamefont {Foss-Feig}\ \emph {et~al.}(2017)\citenamefont
		{Foss-Feig}, \citenamefont {Niroula}, \citenamefont {Young}, \citenamefont
		{Hafezi}, \citenamefont {Gorshkov}, \citenamefont {Wilson},\ and\
		\citenamefont {Maghrebi}}]{Foss-FeigPRA17}%
	\BibitemOpen
	\bibinfo {author} {M.~Foss-Feig}, \bibinfo {author} {P.~Niroula}, \bibinfo
	{author} {J.~T. Young}, \bibinfo {author} {M.~Hafezi}, \bibinfo {author}
	{A.~V. Gorshkov}, \bibinfo {author} {R.~M. Wilson},\ and\ \bibinfo {author}
	{M.~F. Maghrebi},\ \emph {\bibinfo {title} {Emergent equilibrium in many-body
			optical bistability}},\ \href
	{https://link.aps.org/doi/10.1103/PhysRevA.95.043826} {\bibfield  {journal}
		{\bibinfo  {journal} {Phys. Rev. A}\ }\textbf {\bibinfo {volume} {95}},\
		\bibinfo {pages} {043826} (\bibinfo {year} {2017})}\BibitemShut {NoStop}%
	\bibitem [{\citenamefont {Biella}\ \emph {et~al.}(2017)\citenamefont {Biella},
		\citenamefont {Storme}, \citenamefont {Lebreuilly}, \citenamefont {Rossini},
		\citenamefont {Fazio}, \citenamefont {Carusotto},\ and\ \citenamefont
		{Ciuti}}]{BiellaPRA17}%
	\BibitemOpen
	\bibinfo {author} {A.~Biella}, \bibinfo {author} {F.~Storme}, \bibinfo
	{author} {J.~Lebreuilly}, \bibinfo {author} {D.~Rossini}, \bibinfo {author}
	{R.~Fazio}, \bibinfo {author} {I.~Carusotto},\ and\ \bibinfo {author}
	{C.~Ciuti},\ \emph {\bibinfo {title} {Phase diagram of incoherently driven
			strongly correlated photonic lattices}},\ \href
	{https://link.aps.org/doi/10.1103/PhysRevA.96.023839} {\bibfield  {journal}
		{\bibinfo  {journal} {Phys. Rev. A}\ }\textbf {\bibinfo {volume} {96}},\
		\bibinfo {pages} {023839} (\bibinfo {year} {2017})}\BibitemShut {NoStop}%
	\bibitem [{\citenamefont {Savona}(2017)}]{SavonaPRA17}%
	\BibitemOpen
	\bibinfo {author} {V.~Savona},\ \emph {\bibinfo {title} {Spontaneous symmetry
			breaking in a quadratically driven nonlinear photonic lattice}},\ \href
	{https://link.aps.org/doi/10.1103/PhysRevA.96.033826} {\bibfield  {journal}
		{\bibinfo  {journal} {Phys. Rev. A}\ }\textbf {\bibinfo {volume} {96}},\
		\bibinfo {pages} {033826} (\bibinfo {year} {2017})}\BibitemShut {NoStop}%
	\bibitem [{\citenamefont {Lee}\ \emph {et~al.}(2013)\citenamefont {Lee},
		\citenamefont {Gopalakrishnan},\ and\ \citenamefont {Lukin}}]{LeePRL13}%
	\BibitemOpen
	\bibinfo {author} {T.~E. Lee}, \bibinfo {author} {S.~Gopalakrishnan},\ and\
	\bibinfo {author} {M.~D. Lukin},\ \emph {\bibinfo {title} {Unconventional
			Magnetism via Optical Pumping of Interacting Spin Systems}},\ \href
	{http://link.aps.org/doi/10.1103/PhysRevLett.110.257204} {\bibfield
		{journal} {\bibinfo  {journal} {Phys. Rev. Lett.}\ }\textbf {\bibinfo
			{volume} {110}},\ \bibinfo {pages} {257204} (\bibinfo {year}
		{2013})}\BibitemShut {NoStop}%
	\bibitem [{\citenamefont {Jin}\ \emph {et~al.}(2016)\citenamefont {Jin},
		\citenamefont {Biella}, \citenamefont {Viyuela}, \citenamefont {Mazza},
		\citenamefont {Keeling}, \citenamefont {Fazio},\ and\ \citenamefont
		{Rossini}}]{JinPRX16}%
	\BibitemOpen
	\bibinfo {author} {J.~Jin}, \bibinfo {author} {A.~Biella}, \bibinfo {author}
	{O.~Viyuela}, \bibinfo {author} {L.~Mazza}, \bibinfo {author} {J.~Keeling},
	\bibinfo {author} {R.~Fazio},\ and\ \bibinfo {author} {D.~Rossini},\ \emph
	{\bibinfo {title} {Cluster Mean-Field Approach to the Steady-State Phase
			Diagram of Dissipative Spin Systems}},\ \href
	{http://link.aps.org/doi/10.1103/PhysRevX.6.031011} {\bibfield  {journal}
		{\bibinfo  {journal} {Phys. Rev. X}\ }\textbf {\bibinfo {volume} {6}},\
		\bibinfo {pages} {031011} (\bibinfo {year} {2016})}\BibitemShut {NoStop}%
	\bibitem [{\citenamefont {Rota}\ \emph {et~al.}(2017)\citenamefont {Rota},
		\citenamefont {Storme}, \citenamefont {Bartolo}, \citenamefont {Fazio},\ and\
		\citenamefont {Ciuti}}]{RotaPRB17}%
	\BibitemOpen
	\bibinfo {author} {R.~Rota}, \bibinfo {author} {F.~Storme}, \bibinfo {author}
	{N.~Bartolo}, \bibinfo {author} {R.~Fazio},\ and\ \bibinfo {author}
	{C.~Ciuti},\ \emph {\bibinfo {title} {Critical behavior of dissipative
			two-dimensional spin lattices}},\ \href
	{https://link.aps.org/doi/10.1103/PhysRevB.95.134431} {\bibfield  {journal}
		{\bibinfo  {journal} {Phys. Rev. B}\ }\textbf {\bibinfo {volume} {95}},\
		\bibinfo {pages} {134431} (\bibinfo {year} {2017})}\BibitemShut {NoStop}%
	\bibitem [{\citenamefont {Hwang}\ \emph {et~al.}(2018)\citenamefont {Hwang},
		\citenamefont {Rabl},\ and\ \citenamefont {Plenio}}]{HwangPRA18}%
	\BibitemOpen
	\bibinfo {author} {M.-J. Hwang}, \bibinfo {author} {P.~Rabl},\ and\ \bibinfo
	{author} {M.~B. Plenio},\ \emph {\bibinfo {title} {Dissipative phase
			transition in the open quantum Rabi model}},\ \href {\doibase
		10.1103/PhysRevA.97.013825} {\bibfield  {journal} {\bibinfo  {journal} {Phys.
				Rev. A}\ }\textbf {\bibinfo {volume} {97}},\ \bibinfo {pages} {013825}
		(\bibinfo {year} {2018})}\BibitemShut {NoStop}%
	\bibitem [{\citenamefont {Rota}\ \emph {et~al.}(2018)\citenamefont {Rota},
		\citenamefont {Minganti}, \citenamefont {Biella},\ and\ \citenamefont
		{Ciuti}}]{RotaNJP18}%
	\BibitemOpen
	\bibinfo {author} {R.~Rota}, \bibinfo {author} {F.~Minganti}, \bibinfo
	{author} {A.~Biella},\ and\ \bibinfo {author} {C.~Ciuti},\ \emph {\bibinfo
		{title} {Dynamical properties of dissipative XYZ Heisenberg lattices}},\
	\href {http://stacks.iop.org/1367-2630/20/i=4/a=045003} {\bibfield  {journal}
		{\bibinfo  {journal} {New J. Phys.}\ }\textbf {\bibinfo {volume} {20}},\
		\bibinfo {pages} {045003} (\bibinfo {year} {2018})}\BibitemShut {NoStop}%
	\bibitem [{\citenamefont {Rota}\ \emph {et~al.}(2019)\citenamefont {Rota},
		\citenamefont {Minganti}, \citenamefont {Ciuti},\ and\ \citenamefont
		{Savona}}]{RotaPRL19}%
	\BibitemOpen
	\bibinfo {author} {R.~Rota}, \bibinfo {author} {F.~Minganti}, \bibinfo
	{author} {C.~Ciuti},\ and\ \bibinfo {author} {V.~Savona},\ \emph {\bibinfo
		{title} {Quantum Critical Regime in a Quadratically Driven Nonlinear Photonic
			Lattice}},\ \href {https://link.aps.org/doi/10.1103/PhysRevLett.122.110405}
	{\bibfield  {journal} {\bibinfo  {journal} {Phys. Rev. Lett.}\ }\textbf
		{\bibinfo {volume} {122}},\ \bibinfo {pages} {110405} (\bibinfo {year}
		{2019})}\BibitemShut {NoStop}%
	\bibitem [{\citenamefont {Huybrechts}\ \emph {et~al.}(2020)\citenamefont
		{Huybrechts}, \citenamefont {Minganti}, \citenamefont {Nori}, \citenamefont
		{Wouters},\ and\ \citenamefont {Shammah}}]{HuybrechtsPRB20}%
	\BibitemOpen
	\bibinfo {author} {D.~Huybrechts}, \bibinfo {author} {F.~Minganti}, \bibinfo
	{author} {F.~Nori}, \bibinfo {author} {M.~Wouters},\ and\ \bibinfo {author}
	{N.~Shammah},\ \emph {\bibinfo {title} {Validity of mean-field theory in a
			dissipative critical system: Liouvillian gap, $\mathbb{PT}$-symmetric
			antigap, and permutational symmetry in the $XYZ$ model}},\ \href {\doibase
		10.1103/PhysRevB.101.214302} {\bibfield  {journal} {\bibinfo  {journal}
			{Phys. Rev. B}\ }\textbf {\bibinfo {volume} {101}},\ \bibinfo {pages}
		{214302} (\bibinfo {year} {2020})}\BibitemShut {NoStop}%
	\bibitem [{\citenamefont {Garbe}\ \emph {et~al.}(2020)\citenamefont {Garbe},
		\citenamefont {Bina}, \citenamefont {Keller}, \citenamefont {Paris},\ and\
		\citenamefont {Felicetti}}]{GarbePRL20}%
	\BibitemOpen
	\bibinfo {author} {L.~Garbe}, \bibinfo {author} {M.~Bina}, \bibinfo {author}
	{A.~Keller}, \bibinfo {author} {M.~G.~A. Paris},\ and\ \bibinfo {author}
	{S.~Felicetti},\ \emph {\bibinfo {title} {Critical Quantum Metrology with a
			Finite-Component Quantum Phase Transition}},\ \href {\doibase
		10.1103/PhysRevLett.124.120504} {\bibfield  {journal} {\bibinfo  {journal}
			{Phys. Rev. Lett.}\ }\textbf {\bibinfo {volume} {124}},\ \bibinfo {pages}
		{120504} (\bibinfo {year} {2020})}\BibitemShut {NoStop}%
	\bibitem [{\citenamefont {Landa}\ \emph
		{et~al.}(2020{\natexlab{a}})\citenamefont {Landa}, \citenamefont {Schir\'o},\
		and\ \citenamefont {Misguich}}]{LandaPRL20}%
	\BibitemOpen
	\bibinfo {author} {H.~Landa}, \bibinfo {author} {M.~Schir\'o},\ and\ \bibinfo
	{author} {G.~Misguich},\ \emph {\bibinfo {title} {Multistability of
			Driven-Dissipative Quantum Spins}},\ \href {\doibase
		10.1103/PhysRevLett.124.043601} {\bibfield  {journal} {\bibinfo  {journal}
			{Phys. Rev. Lett.}\ }\textbf {\bibinfo {volume} {124}},\ \bibinfo {pages}
		{043601} (\bibinfo {year} {2020}{\natexlab{a}})}\BibitemShut {NoStop}%
	\bibitem [{\citenamefont {Curtis}\ \emph {et~al.}(2020)\citenamefont {Curtis},
		\citenamefont {Boettcher}, \citenamefont {Young}, \citenamefont {Maghrebi},
		\citenamefont {Carmichael}, \citenamefont {Gorshkov},\ and\ \citenamefont
		{Foss-Feig}}]{CurtisarXiv20}%
	\BibitemOpen
	\bibinfo {author} {J.~B. Curtis}, \bibinfo {author} {I.~Boettcher}, \bibinfo
	{author} {J.~T. Young}, \bibinfo {author} {M.~F. Maghrebi}, \bibinfo {author}
	{H.~Carmichael}, \bibinfo {author} {A.~V. Gorshkov},\ and\ \bibinfo {author}
	{M.~Foss-Feig},\ \href@noop {} {\emph {\bibinfo {title} {Critical Theory for
				the Breakdown of Photon Blockade}}} (\bibinfo {year} {2020}),\ \Eprint
	{http://arxiv.org/abs/arXiv:2006.05593} {arXiv:2006.05593} \BibitemShut
	{NoStop}%
	\bibitem [{\citenamefont {Fitzpatrick}\ \emph {et~al.}(2017)\citenamefont
		{Fitzpatrick}, \citenamefont {Sundaresan}, \citenamefont {Li}, \citenamefont
		{Koch},\ and\ \citenamefont {Houck}}]{FitzpatrickPRX17}%
	\BibitemOpen
	\bibinfo {author} {M.~Fitzpatrick}, \bibinfo {author} {N.~M. Sundaresan},
	\bibinfo {author} {A.~C.~Y. Li}, \bibinfo {author} {J.~Koch},\ and\ \bibinfo
	{author} {A.~A. Houck},\ \emph {\bibinfo {title} {Observation of a
			Dissipative Phase Transition in a One-Dimensional Circuit QED Lattice}},\
	\href {https://link.aps.org/doi/10.1103/PhysRevX.7.011016} {\bibfield
		{journal} {\bibinfo  {journal} {Phys. Rev. X}\ }\textbf {\bibinfo {volume}
			{7}},\ \bibinfo {pages} {011016} (\bibinfo {year} {2017})}\BibitemShut
	{NoStop}%
	\bibitem [{\citenamefont {M\"uller}\ \emph {et~al.}(2012)\citenamefont
		{M\"uller}, \citenamefont {Diehl}, \citenamefont {Pupillo},\ and\
		\citenamefont {Zoller}}]{Mueller_2012}%
	\BibitemOpen
	\bibinfo {author} {M.~M\"uller}, \bibinfo {author} {S.~Diehl}, \bibinfo
	{author} {G.~Pupillo},\ and\ \bibinfo {author} {P.~Zoller},\ \emph {\bibinfo
		{title} {Engineered Open Systems and Quantum Simulations with Atoms and
			Ions}},\ \href
	{https://www.sciencedirect.com/science/article/pii/B9780123964823000016}
	{\bibfield  {journal} {\bibinfo  {journal} {Adv. At. Mol. Opt. Phys.}\
		}\textbf {\bibinfo {volume} {61}},\ \bibinfo {pages} {1} (\bibinfo {year}
		{2012})}\BibitemShut {NoStop}%
	\bibitem [{\citenamefont {Bernien}\ \emph {et~al.}(2017)\citenamefont
		{Bernien}, \citenamefont {Schwartz}, \citenamefont {Keesling}, \citenamefont
		{Levine}, \citenamefont {Omran}, \citenamefont {Pichler}, \citenamefont
		{Choi}, \citenamefont {Zibrov}, \citenamefont {Endres}, \citenamefont
		{Greiner}, \citenamefont {Vuleti{\'c}},\ and\ \citenamefont
		{Lukin}}]{BernienNAT2017}%
	\BibitemOpen
	\bibinfo {author} {H.~Bernien}, \bibinfo {author} {S.~Schwartz}, \bibinfo
	{author} {A.~Keesling}, \bibinfo {author} {H.~Levine}, \bibinfo {author}
	{A.~Omran}, \bibinfo {author} {H.~Pichler}, \bibinfo {author} {S.~Choi},
	\bibinfo {author} {A.~S. Zibrov}, \bibinfo {author} {M.~Endres}, \bibinfo
	{author} {M.~Greiner}, \bibinfo {author} {V.~Vuleti{\'c}},\ and\ \bibinfo
	{author} {M.~D. Lukin},\ \emph {\bibinfo {title} {Probing many-body dynamics
			on a 51-atom quantum simulator}},\ \href
	{http://dx.doi.org/10.1038/nature24622} {\bibfield  {journal} {\bibinfo
			{journal} {Nature (London)}\ }\textbf {\bibinfo {volume} {551}},\ \bibinfo
		{pages} {579} (\bibinfo {year} {2017})}\BibitemShut {NoStop}%
	\bibitem [{\citenamefont {Aspelmeyer}\ \emph {et~al.}(2014)\citenamefont
		{Aspelmeyer}, \citenamefont {Kippenberg},\ and\ \citenamefont
		{Marquardt}}]{AspelmeyerRMP14}%
	\BibitemOpen
	\bibinfo {author} {M.~Aspelmeyer}, \bibinfo {author} {T.~J. Kippenberg},\
	and\ \bibinfo {author} {F.~Marquardt},\ \emph {\bibinfo {title} {Cavity
			optomechanics}},\ \href {https://link.aps.org/doi/10.1103/RevModPhys.86.1391}
	{\bibfield  {journal} {\bibinfo  {journal} {Rev. Mod. Phys.}\ }\textbf
		{\bibinfo {volume} {86}},\ \bibinfo {pages} {1391} (\bibinfo {year}
		{2014})}\BibitemShut {NoStop}%
	\bibitem [{\citenamefont {Gil-Santos}\ \emph {et~al.}(2017)\citenamefont
		{Gil-Santos}, \citenamefont {Labousse}, \citenamefont {Baker}, \citenamefont
		{Goetschy}, \citenamefont {Hease}, \citenamefont {Gomez}, \citenamefont
		{Lema\^{\i}tre}, \citenamefont {Leo}, \citenamefont {Ciuti},\ and\
		\citenamefont {Favero}}]{GilSantosPRL17}%
	\BibitemOpen
	\bibinfo {author} {E.~Gil-Santos}, \bibinfo {author} {M.~Labousse}, \bibinfo
	{author} {C.~Baker}, \bibinfo {author} {A.~Goetschy}, \bibinfo {author}
	{W.~Hease}, \bibinfo {author} {C.~Gomez}, \bibinfo {author}
	{A.~Lema\^{\i}tre}, \bibinfo {author} {G.~Leo}, \bibinfo {author}
	{C.~Ciuti},\ and\ \bibinfo {author} {I.~Favero},\ \emph {\bibinfo {title}
		{Light-Mediated Cascaded Locking of Multiple Nano-Optomechanical
			Oscillators}},\ \href
	{https://link.aps.org/doi/10.1103/PhysRevLett.118.063605} {\bibfield
		{journal} {\bibinfo  {journal} {Phys. Rev. Lett.}\ }\textbf {\bibinfo
			{volume} {118}},\ \bibinfo {pages} {063605} (\bibinfo {year}
		{2017})}\BibitemShut {NoStop}%
	\bibitem [{\citenamefont {Kasprzak}\ \emph {et~al.}(2006)\citenamefont
		{Kasprzak}, \citenamefont {Richard}, \citenamefont {Kundermann},
		\citenamefont {Baas}, \citenamefont {Jeambrun}, \citenamefont {Keeling},
		\citenamefont {Marchetti}, \citenamefont {Szymanska}, \citenamefont {Andre},
		\citenamefont {Staehli}, \citenamefont {Savona}, \citenamefont {Littlewood},
		\citenamefont {Deveaud},\ and\ \citenamefont {Dang}}]{KasprzakNAT2006}%
	\BibitemOpen
	\bibinfo {author} {J.~Kasprzak}, \bibinfo {author} {M.~Richard}, \bibinfo
	{author} {S.~Kundermann}, \bibinfo {author} {A.~Baas}, \bibinfo {author}
	{P.~Jeambrun}, \bibinfo {author} {J.~M.~J. Keeling}, \bibinfo {author} {F.~M.
		Marchetti}, \bibinfo {author} {M.~H. Szymanska}, \bibinfo {author}
	{R.~Andre}, \bibinfo {author} {J.~L. Staehli}, \bibinfo {author} {V.~Savona},
	\bibinfo {author} {P.~B. Littlewood}, \bibinfo {author} {B.~Deveaud},\ and\
	\bibinfo {author} {L.~S. Dang},\ \emph {\bibinfo {title} {{Bose-Einstein}
			condensation of exciton polaritons}},\ \href
	{http://dx.doi.org/10.1038/nature05131} {\bibfield  {journal} {\bibinfo
			{journal} {Nature (London)}\ }\textbf {\bibinfo {volume} {443}},\ \bibinfo
		{pages} {409} (\bibinfo {year} {2006})}\BibitemShut {NoStop}%
	\bibitem [{\citenamefont {Fink}\ \emph {et~al.}(2017)\citenamefont {Fink},
		\citenamefont {Dombi}, \citenamefont {Vukics}, \citenamefont {Wallraff},\
		and\ \citenamefont {Domokos}}]{FinkPRX17}%
	\BibitemOpen
	\bibinfo {author} {J.~M. Fink}, \bibinfo {author} {A.~Dombi}, \bibinfo
	{author} {A.~Vukics}, \bibinfo {author} {A.~Wallraff},\ and\ \bibinfo
	{author} {P.~Domokos},\ \emph {\bibinfo {title} {Observation of the
			Photon-Blockade Breakdown Phase Transition}},\ \href
	{https://link.aps.org/doi/10.1103/PhysRevX.7.011012} {\bibfield  {journal}
		{\bibinfo  {journal} {Phys. Rev. X}\ }\textbf {\bibinfo {volume} {7}},\
		\bibinfo {pages} {011012} (\bibinfo {year} {2017})}\BibitemShut {NoStop}%
	\bibitem [{\citenamefont {Rodriguez}\ \emph {et~al.}(2017)\citenamefont
		{Rodriguez}, \citenamefont {Casteels}, \citenamefont {Storme}, \citenamefont
		{Carlon~Zambon}, \citenamefont {Sagnes}, \citenamefont {Le~Gratiet},
		\citenamefont {Galopin}, \citenamefont {Lema\^{\i}tre}, \citenamefont {Amo},
		\citenamefont {Ciuti},\ and\ \citenamefont {Bloch}}]{RodriguezPRL17}%
	\BibitemOpen
	\bibinfo {author} {S.~R.~K. Rodriguez}, \bibinfo {author} {W.~Casteels},
	\bibinfo {author} {F.~Storme}, \bibinfo {author} {N.~Carlon~Zambon}, \bibinfo
	{author} {I.~Sagnes}, \bibinfo {author} {L.~Le~Gratiet}, \bibinfo {author}
	{E.~Galopin}, \bibinfo {author} {A.~Lema\^{\i}tre}, \bibinfo {author}
	{A.~Amo}, \bibinfo {author} {C.~Ciuti},\ and\ \bibinfo {author} {J.~Bloch},\
	\emph {\bibinfo {title} {Probing a Dissipative Phase Transition via Dynamical
			Optical Hysteresis}},\ \href
	{https://link.aps.org/doi/10.1103/PhysRevLett.118.247402} {\bibfield
		{journal} {\bibinfo  {journal} {Phys. Rev. Lett.}\ }\textbf {\bibinfo
			{volume} {118}},\ \bibinfo {pages} {247402} (\bibinfo {year}
		{2017})}\BibitemShut {NoStop}%
	\bibitem [{\citenamefont {Shammah}\ \emph {et~al.}(2018)\citenamefont
		{Shammah}, \citenamefont {Ahmed}, \citenamefont {Lambert}, \citenamefont
		{De~Liberato},\ and\ \citenamefont {Nori}}]{ShammahPRA18}%
	\BibitemOpen
	\bibinfo {author} {N.~Shammah}, \bibinfo {author} {S.~Ahmed}, \bibinfo
	{author} {N.~Lambert}, \bibinfo {author} {S.~De~Liberato},\ and\ \bibinfo
	{author} {F.~Nori},\ \emph {\bibinfo {title} {Open quantum systems with local
			and collective incoherent processes: Efficient numerical simulations using
			permutational invariance}},\ \href {\doibase 10.1103/PhysRevA.98.063815}
	{\bibfield  {journal} {\bibinfo  {journal} {Phys. Rev. A}\ }\textbf {\bibinfo
			{volume} {98}},\ \bibinfo {pages} {063815} (\bibinfo {year}
		{2018})}\BibitemShut {NoStop}%
	\bibitem [{\citenamefont {Tucker}\ \emph {et~al.}(2018)\citenamefont {Tucker},
		\citenamefont {Zhu}, \citenamefont {Lewis-Swan}, \citenamefont {Marino},
		\citenamefont {Jimenez}, \citenamefont {Restrepo},\ and\ \citenamefont
		{Rey}}]{TuckerNJP18}%
	\BibitemOpen
	\bibinfo {author} {K.~Tucker}, \bibinfo {author} {B.~Zhu}, \bibinfo {author}
	{R.~J. Lewis-Swan}, \bibinfo {author} {J.~Marino}, \bibinfo {author}
	{F.~Jimenez}, \bibinfo {author} {J.~G. Restrepo},\ and\ \bibinfo {author}
	{A.~M. Rey},\ \emph {\bibinfo {title} {Shattered time: can a dissipative time
			crystal survive many-body correlations?}},\ \href {\doibase
		10.1088/1367-2630/aaf18b} {\bibfield  {journal} {\bibinfo  {journal} {New J.
				Phys.}\ }\textbf {\bibinfo {volume} {20}},\ \bibinfo {pages} {123003}
		(\bibinfo {year} {2018})}\BibitemShut {NoStop}%
	\bibitem [{\citenamefont {Lled\'o}\ \emph {et~al.}(2019)\citenamefont
		{Lled\'o}, \citenamefont {Mavrogordatos},\ and\ \citenamefont
		{Szyma\'{n}ska}}]{LLedoPRB19}%
	\BibitemOpen
	\bibinfo {author} {C.~Lled\'o}, \bibinfo {author} {T.~K. Mavrogordatos},\
	and\ \bibinfo {author} {M.~H. Szyma\'{n}ska},\ \emph {\bibinfo {title}
		{Driven Bose-Hubbard dimer under nonlocal dissipation: A bistable time
			crystal}},\ \href {\doibase 10.1103/PhysRevB.100.054303} {\bibfield
		{journal} {\bibinfo  {journal} {Phys. Rev. B}\ }\textbf {\bibinfo {volume}
			{100}},\ \bibinfo {pages} {054303} (\bibinfo {year} {2019})}\BibitemShut
	{NoStop}%
	\bibitem [{\citenamefont {Bu{\v{c}}a}\ \emph {et~al.}(2019)\citenamefont
		{Bu{\v{c}}a}, \citenamefont {Tindall},\ and\ \citenamefont
		{Jaksch}}]{BucaNat19}%
	\BibitemOpen
	\bibinfo {author} {B.~Bu{\v{c}}a}, \bibinfo {author} {J.~Tindall},\ and\
	\bibinfo {author} {D.~Jaksch},\ \emph {\bibinfo {title} {Non-stationary
			coherent quantum many-body dynamics through dissipation}},\ \href {\doibase
		10.1038/s41467-019-09757-y} {\bibfield  {journal} {\bibinfo  {journal} {Nat.
				Commun.}\ }\textbf {\bibinfo {volume} {10}},\ \bibinfo {pages} {1730}
		(\bibinfo {year} {2019})}\BibitemShut {NoStop}%
	\bibitem [{\citenamefont {Seibold}\ \emph {et~al.}(2020)\citenamefont
		{Seibold}, \citenamefont {Rota},\ and\ \citenamefont
		{Savona}}]{SeiboldPRA20}%
	\BibitemOpen
	\bibinfo {author} {K.~Seibold}, \bibinfo {author} {R.~Rota},\ and\ \bibinfo
	{author} {V.~Savona},\ \emph {\bibinfo {title} {Dissipative time crystal in
			an asymmetric nonlinear photonic dimer}},\ \href {\doibase
		10.1103/PhysRevA.101.033839} {\bibfield  {journal} {\bibinfo  {journal}
			{Phys. Rev. A}\ }\textbf {\bibinfo {volume} {101}},\ \bibinfo {pages}
		{033839} (\bibinfo {year} {2020})}\BibitemShut {NoStop}%
	\bibitem [{\citenamefont {Sacha}\ and\ \citenamefont
		{Zakrzewski}(2017)}]{SachaRPP17}%
	\BibitemOpen
	\bibinfo {author} {K.~Sacha}\ and\ \bibinfo {author} {J.~Zakrzewski},\ \emph
	{\bibinfo {title} {Time crystals: a review}},\ \href {\doibase
		10.1088/1361-6633/aa8b38} {\bibfield  {journal} {\bibinfo  {journal} {Rep.
				Prog. Phys.}\ }\textbf {\bibinfo {volume} {81}},\ \bibinfo {pages} {016401}
		(\bibinfo {year} {2017})}\BibitemShut {NoStop}%
	\bibitem [{\citenamefont {Gong}\ \emph {et~al.}(2018)\citenamefont {Gong},
		\citenamefont {Hamazaki},\ and\ \citenamefont {Ueda}}]{GongPRL18}%
	\BibitemOpen
	\bibinfo {author} {Z.~Gong}, \bibinfo {author} {R.~Hamazaki},\ and\ \bibinfo
	{author} {M.~Ueda},\ \emph {\bibinfo {title} {Discrete Time-Crystalline Order
			in Cavity and Circuit QED Systems}},\ \href {\doibase
		10.1103/PhysRevLett.120.040404} {\bibfield  {journal} {\bibinfo  {journal}
			{Phys. Rev. Lett.}\ }\textbf {\bibinfo {volume} {120}},\ \bibinfo {pages}
		{040404} (\bibinfo {year} {2018})}\BibitemShut {NoStop}%
	\bibitem [{\citenamefont {Riera-Campeny}\ \emph {et~al.}(2020)\citenamefont
		{Riera-Campeny}, \citenamefont {Moreno-Cardoner},\ and\ \citenamefont
		{Sanpera}}]{RieraQuantum2020}%
	\BibitemOpen
	\bibinfo {author} {A.~Riera-Campeny}, \bibinfo {author}
	{M.~Moreno-Cardoner},\ and\ \bibinfo {author} {A.~Sanpera},\ \emph {\bibinfo
		{title} {Time crystallinity in open quantum systems}},\ \href {\doibase
		10.22331/q-2020-05-25-270} {\bibfield  {journal} {\bibinfo  {journal}
			{{Quantum}}\ }\textbf {\bibinfo {volume} {4}},\ \bibinfo {pages} {270}
		(\bibinfo {year} {2020})}\BibitemShut {NoStop}%
	\bibitem [{\citenamefont {Gambetta}\ \emph {et~al.}(2019)\citenamefont
		{Gambetta}, \citenamefont {Carollo}, \citenamefont {Marcuzzi}, \citenamefont
		{Garrahan},\ and\ \citenamefont {Lesanovsky}}]{GambettaPRL19}%
	\BibitemOpen
	\bibinfo {author} {F.~M. Gambetta}, \bibinfo {author} {F.~Carollo}, \bibinfo
	{author} {M.~Marcuzzi}, \bibinfo {author} {J.~P. Garrahan},\ and\ \bibinfo
	{author} {I.~Lesanovsky},\ \emph {\bibinfo {title} {Discrete Time Crystals in
			the Absence of Manifest Symmetries or Disorder in Open Quantum Systems}},\
	\href {\doibase 10.1103/PhysRevLett.122.015701} {\bibfield  {journal}
		{\bibinfo  {journal} {Phys. Rev. Lett.}\ }\textbf {\bibinfo {volume} {122}},\
		\bibinfo {pages} {015701} (\bibinfo {year} {2019})}\BibitemShut {NoStop}%
	\bibitem [{\citenamefont {Scarlatella}\ \emph {et~al.}(2019)\citenamefont
		{Scarlatella}, \citenamefont {Fazio},\ and\ \citenamefont
		{Schir\'o}}]{ScarlatellaPRB19}%
	\BibitemOpen
	\bibinfo {author} {O.~Scarlatella}, \bibinfo {author} {R.~Fazio},\ and\
	\bibinfo {author} {M.~Schir\'o},\ \emph {\bibinfo {title} {Emergent finite
			frequency criticality of driven-dissipative correlated lattice bosons}},\
	\href {\doibase 10.1103/PhysRevB.99.064511} {\bibfield  {journal} {\bibinfo
			{journal} {Phys. Rev. B}\ }\textbf {\bibinfo {volume} {99}},\ \bibinfo
		{pages} {064511} (\bibinfo {year} {2019})}\BibitemShut {NoStop}%
	\bibitem [{\citenamefont {Kessler}\ \emph {et~al.}(2012)\citenamefont
		{Kessler}, \citenamefont {Giedke}, \citenamefont {Imamoglu}, \citenamefont
		{Yelin}, \citenamefont {Lukin},\ and\ \citenamefont {Cirac}}]{KesslerPRA12}%
	\BibitemOpen
	\bibinfo {author} {E.~M. Kessler}, \bibinfo {author} {G.~Giedke}, \bibinfo
	{author} {A.~Imamoglu}, \bibinfo {author} {S.~F. Yelin}, \bibinfo {author}
	{M.~D. Lukin},\ and\ \bibinfo {author} {J.~I. Cirac},\ \emph {\bibinfo
		{title} {Dissipative phase transition in a central spin system}},\ \href
	{https://link.aps.org/doi/10.1103/PhysRevA.86.012116} {\bibfield  {journal}
		{\bibinfo  {journal} {Phys. Rev. A}\ }\textbf {\bibinfo {volume} {86}},\
		\bibinfo {pages} {012116} (\bibinfo {year} {2012})}\BibitemShut {NoStop}%
	\bibitem [{\citenamefont {Booker}\ \emph {et~al.}(2020)\citenamefont {Booker},
		\citenamefont {Bu\v{c}a},\ and\ \citenamefont {Jaksch}}]{Bookerarxiv20}%
	\BibitemOpen
	\bibinfo {author} {C.~Booker}, \bibinfo {author} {B.~Bu\v{c}a},\ and\
	\bibinfo {author} {D.~Jaksch},\ \emph {\bibinfo {title} {Non-stationarity and
			Dissipative Time Crystals: Spectral Properties and Finite-Size Effects}},\
	\href@noop {} {\  (\bibinfo {year} {2020})},\ \Eprint
	{http://arxiv.org/abs/arXiv:2005.05062} {arXiv:2005.05062} \BibitemShut
	{NoStop}%
	\bibitem [{\citenamefont {Lebreuilly}\ \emph {et~al.}(2017)\citenamefont
		{Lebreuilly}, \citenamefont {Biella}, \citenamefont {Storme}, \citenamefont
		{Rossini}, \citenamefont {Fazio}, \citenamefont {Ciuti},\ and\ \citenamefont
		{Carusotto}}]{LebreuillyPRA17}%
	\BibitemOpen
	\bibinfo {author} {J.~Lebreuilly}, \bibinfo {author} {A.~Biella}, \bibinfo
	{author} {F.~Storme}, \bibinfo {author} {D.~Rossini}, \bibinfo {author}
	{R.~Fazio}, \bibinfo {author} {C.~Ciuti},\ and\ \bibinfo {author}
	{I.~Carusotto},\ \emph {\bibinfo {title} {Stabilizing strongly correlated
			photon fluids with non-Markovian reservoirs}},\ \href
	{https://link.aps.org/doi/10.1103/PhysRevA.96.033828} {\bibfield  {journal}
		{\bibinfo  {journal} {Phys. Rev. A}\ }\textbf {\bibinfo {volume} {96}},\
		\bibinfo {pages} {033828} (\bibinfo {year} {2017})}\BibitemShut {NoStop}%
	\bibitem [{\citenamefont {Takemura}\ \emph {et~al.}(2019)\citenamefont
		{Takemura}, \citenamefont {Takiguchi},\ and\ \citenamefont
		{Notomi}}]{TakemuraArXiv19}%
	\BibitemOpen
	\bibinfo {author} {N.~Takemura}, \bibinfo {author} {M.~Takiguchi},\ and\
	\bibinfo {author} {M.~Notomi},\ \emph {\bibinfo {title} {Low- and
			high-$\beta$ lasers in Class-A limit: photon statistics, linewidth, and the
			laser-phase transition analogy}},\ \href@noop {} {\  (\bibinfo {year}
		{2019})},\ \Eprint {http://arxiv.org/abs/arXiv:1904.01743} {arXiv:1904.01743}
	\BibitemShut {NoStop}%
	\bibitem [{\citenamefont {Leghtas}\ \emph {et~al.}(2015)\citenamefont
		{Leghtas}, \citenamefont {Touzard}, \citenamefont {Pop}, \citenamefont {Kou},
		\citenamefont {Vlastakis}, \citenamefont {Petrenko}, \citenamefont {Sliwa},
		\citenamefont {Narla}, \citenamefont {Shankar}, \citenamefont {Hatridge},
		\citenamefont {Reagor}, \citenamefont {Frunzio}, \citenamefont {Schoelkopf},
		\citenamefont {Mirrahimi},\ and\ \citenamefont {Devoret}}]{LeghtasScience15}%
	\BibitemOpen
	\bibinfo {author} {Z.~Leghtas}, \bibinfo {author} {S.~Touzard}, \bibinfo
	{author} {I.~M. Pop}, \bibinfo {author} {A.~Kou}, \bibinfo {author}
	{B.~Vlastakis}, \bibinfo {author} {A.~Petrenko}, \bibinfo {author} {K.~M.
		Sliwa}, \bibinfo {author} {A.~Narla}, \bibinfo {author} {S.~Shankar},
	\bibinfo {author} {M.~J. Hatridge}, \bibinfo {author} {M.~Reagor}, \bibinfo
	{author} {L.~Frunzio}, \bibinfo {author} {R.~J. Schoelkopf}, \bibinfo
	{author} {M.~Mirrahimi},\ and\ \bibinfo {author} {M.~H. Devoret},\ \emph
	{\bibinfo {title} {Confining the state of light to a quantum manifold by
			engineered two-photon loss}},\ \href
	{http://dx.doi.org/10.1126/science.aaa2085} {\bibfield  {journal} {\bibinfo
			{journal} {Science}\ }\textbf {\bibinfo {volume} {347}},\ \bibinfo {pages}
		{853} (\bibinfo {year} {2015})}\BibitemShut {NoStop}%
	\bibitem [{\citenamefont {Lescanne}\ \emph {et~al.}(2020)\citenamefont
		{Lescanne}, \citenamefont {Villiers}, \citenamefont {Peronnin}, \citenamefont
		{Sarlette}, \citenamefont {Delbecq}, \citenamefont {Huard}, \citenamefont
		{Kontos}, \citenamefont {Mirrahimi},\ and\ \citenamefont
		{Leghtas}}]{LescanneNatPhys2020}%
	\BibitemOpen
	\bibinfo {author} {R.~Lescanne}, \bibinfo {author} {M.~Villiers}, \bibinfo
	{author} {T.~Peronnin}, \bibinfo {author} {A.~Sarlette}, \bibinfo {author}
	{M.~Delbecq}, \bibinfo {author} {B.~Huard}, \bibinfo {author} {T.~Kontos},
	\bibinfo {author} {M.~Mirrahimi},\ and\ \bibinfo {author} {Z.~Leghtas},\
	\emph {\bibinfo {title} {Exponential suppression of bit-flips in a qubit
			encoded in an oscillator}},\ \href {\doibase 10.1038/s41567-020-0824-x}
	{\bibfield  {journal} {\bibinfo  {journal} {Nature Physics}\ }\textbf
		{\bibinfo {volume} {16}},\ \bibinfo {pages} {509} (\bibinfo {year}
		{2020})}\BibitemShut {NoStop}%
	\bibitem [{\citenamefont {Baumgartner}\ and\ \citenamefont
		{Heide}(2008)}]{BaumgartnerJPA08}%
	\BibitemOpen
	\bibinfo {author} {B.~Baumgartner}\ and\ \bibinfo {author} {N.~Heide},\ \emph
	{\bibinfo {title} {Analysis of quantum semigroups with GKS-Lindblad
			generators: II. General}},\ \href
	{http://stacks.iop.org/1751-8121/41/i=39/a=395303} {\bibfield  {journal}
		{\bibinfo  {journal} {J. Phys. A: Math. Theor.}\ }\textbf {\bibinfo {volume}
			{41}},\ \bibinfo {pages} {395303} (\bibinfo {year} {2008})}\BibitemShut
	{NoStop}%
	\bibitem [{\citenamefont {Bu{\v{c}}a}\ and\ \citenamefont
		{Prosen}(2012)}]{BucaNPJ2012}%
	\BibitemOpen
	\bibinfo {author} {B.~Bu{\v{c}}a}\ and\ \bibinfo {author} {T.~Prosen},\ \emph
	{\bibinfo {title} {A note on symmetry reductions of the {L}indblad equation:
			transport in constrained open spin chains}},\ \href {\doibase
		10.1088/1367-2630/14/7/073007} {\bibfield  {journal} {\bibinfo  {journal}
			{New J. Phys.}\ }\textbf {\bibinfo {volume} {14}},\ \bibinfo {pages} {073007}
		(\bibinfo {year} {2012})}\BibitemShut {NoStop}%
	\bibitem [{\citenamefont {Albert}\ and\ \citenamefont
		{Jiang}(2014)}]{AlbertPRA14}%
	\BibitemOpen
	\bibinfo {author} {V.~V. Albert}\ and\ \bibinfo {author} {L.~Jiang},\ \emph
	{\bibinfo {title} {Symmetries and conserved quantities in Lindblad master
			equations}},\ \href {https://link.aps.org/doi/10.1103/PhysRevA.89.022118}
	{\bibfield  {journal} {\bibinfo  {journal} {Phys. Rev. A}\ }\textbf {\bibinfo
			{volume} {89}},\ \bibinfo {pages} {022118} (\bibinfo {year}
		{2014})}\BibitemShut {NoStop}%
	\bibitem [{\citenamefont {Johansson}\ \emph {et~al.}(2012)\citenamefont
		{Johansson}, \citenamefont {Nation},\ and\ \citenamefont {Nori}}]{qutip1}%
	\BibitemOpen
	\bibinfo {author} {J.~Johansson}, \bibinfo {author} {P.~Nation},\ and\
	\bibinfo {author} {F.~Nori},\ \emph {\bibinfo {title} {{QuTiP}: An
			open-source {P}ython framework for the dynamics of open quantum systems}},\
	\href {\doibase 10.1016/j.cpc.2012.02.021} {\bibfield  {journal} {\bibinfo
			{journal} {Comp. Phys. Commun.}\ }\textbf {\bibinfo {volume} {183}},\
		\bibinfo {pages} {1760} (\bibinfo {year} {2012})}\BibitemShut {NoStop}%
	\bibitem [{\citenamefont {Johansson}\ \emph {et~al.}(2013)\citenamefont
		{Johansson}, \citenamefont {Nation},\ and\ \citenamefont {Nori}}]{qutip2}%
	\BibitemOpen
	\bibinfo {author} {J.~Johansson}, \bibinfo {author} {P.~Nation},\ and\
	\bibinfo {author} {F.~Nori},\ \emph {\bibinfo {title} {{QuTiP} 2: A {P}ython
			framework for the dynamics of open quantum systems}},\ \href {\doibase
		10.1016/j.cpc.2012.11.019} {\bibfield  {journal} {\bibinfo  {journal} {Comp.
				Phys. Commun.}\ }\textbf {\bibinfo {volume} {184}},\ \bibinfo {pages} {1234}
		(\bibinfo {year} {2013})}\BibitemShut {NoStop}%
	\bibitem [{\citenamefont {Watanabe}\ and\ \citenamefont
		{Oshikawa}(2015)}]{Watanabe2015}%
	\BibitemOpen
	\bibinfo {author} {H.~Watanabe}\ and\ \bibinfo {author} {M.~Oshikawa},\ \emph
	{\bibinfo {title} {Absence of Quantum Time Crystals}},\ \href {\doibase
		10.1103/PhysRevLett.114.251603} {\bibfield  {journal} {\bibinfo  {journal}
			{Phys. Rev. Lett.}\ }\textbf {\bibinfo {volume} {114}},\ \bibinfo {pages}
		{251603} (\bibinfo {year} {2015})}\BibitemShut {NoStop}%
	\bibitem [{\citenamefont {Sacha}(2015)}]{Sacha2015}%
	\BibitemOpen
	\bibinfo {author} {K.~Sacha},\ \emph {\bibinfo {title} {Modeling spontaneous
			breaking of time-translation symmetry}},\ \href {\doibase
		10.1103/PhysRevA.91.033617} {\bibfield  {journal} {\bibinfo  {journal} {Phys.
				Rev. A}\ }\textbf {\bibinfo {volume} {91}},\ \bibinfo {pages} {033617}
		(\bibinfo {year} {2015})}\BibitemShut {NoStop}%
	\bibitem [{\citenamefont {Choi}\ \emph {et~al.}(2017)\citenamefont {Choi},
		\citenamefont {Choi}, \citenamefont {Landig}, \citenamefont {Kucsko},
		\citenamefont {Zhou}, \citenamefont {Isoya}, \citenamefont {Jelezko},
		\citenamefont {Onoda}, \citenamefont {Sumiya}, \citenamefont {Khemani},
		\citenamefont {von Keyserlingk}, \citenamefont {Yao}, \citenamefont
		{Demler},\ and\ \citenamefont {Lukin}}]{Choi2017}%
	\BibitemOpen
	\bibinfo {author} {S.~Choi}, \bibinfo {author} {J.~Choi}, \bibinfo {author}
	{R.~Landig}, \bibinfo {author} {G.~Kucsko}, \bibinfo {author} {H.~Zhou},
	\bibinfo {author} {J.~Isoya}, \bibinfo {author} {F.~Jelezko}, \bibinfo
	{author} {S.~Onoda}, \bibinfo {author} {H.~Sumiya}, \bibinfo {author}
	{V.~Khemani}, \bibinfo {author} {C.~von Keyserlingk}, \bibinfo {author}
	{N.~Y. Yao}, \bibinfo {author} {E.~Demler},\ and\ \bibinfo {author} {M.~D.
		Lukin},\ \emph {\bibinfo {title} {Observation of discrete time-crystalline
			order in a disordered dipolar many-body system}},\ \href {\doibase
		10.1038/nature21426} {\bibfield  {journal} {\bibinfo  {journal} {Nature
				(London)}\ }\textbf {\bibinfo {volume} {543}},\ \bibinfo {pages} {221}
		(\bibinfo {year} {2017})}\BibitemShut {NoStop}%
	\bibitem [{\citenamefont {Zhang}\ \emph {et~al.}(2017)\citenamefont {Zhang},
		\citenamefont {Hess}, \citenamefont {Kyprianidis}, \citenamefont {Becker},
		\citenamefont {Lee}, \citenamefont {Smith}, \citenamefont {Pagano},
		\citenamefont {Potirniche}, \citenamefont {Potter}, \citenamefont
		{Vishwanath}, \citenamefont {Yao},\ and\ \citenamefont {Monroe}}]{Zhang2017}%
	\BibitemOpen
	\bibinfo {author} {J.~Zhang}, \bibinfo {author} {P.~W. Hess}, \bibinfo
	{author} {A.~Kyprianidis}, \bibinfo {author} {P.~Becker}, \bibinfo {author}
	{A.~Lee}, \bibinfo {author} {J.~Smith}, \bibinfo {author} {G.~Pagano},
	\bibinfo {author} {I.-D. Potirniche}, \bibinfo {author} {A.~C. Potter},
	\bibinfo {author} {A.~Vishwanath}, \bibinfo {author} {N.~Y. Yao},\ and\
	\bibinfo {author} {C.~Monroe},\ \emph {\bibinfo {title} {Observation of a
			discrete time crystal}},\ \href {\doibase 10.1038/nature21413} {\bibfield
		{journal} {\bibinfo  {journal} {Nature (London)}\ }\textbf {\bibinfo {volume}
			{543}},\ \bibinfo {pages} {217} (\bibinfo {year} {2017})}\BibitemShut
	{NoStop}%
	\bibitem [{\citenamefont {Casteels}\ and\ \citenamefont
		{Ciuti}(2017)}]{CasteelsPRA17}%
	\BibitemOpen
	\bibinfo {author} {W.~Casteels}\ and\ \bibinfo {author} {C.~Ciuti},\ \emph
	{\bibinfo {title} {Quantum entanglement in the spatial-symmetry-breaking
			phase transition of a driven-dissipative Bose-Hubbard dimer}},\ \href
	{https://link.aps.org/doi/10.1103/PhysRevA.95.013812} {\bibfield  {journal}
		{\bibinfo  {journal} {Phys. Rev. A}\ }\textbf {\bibinfo {volume} {95}},\
		\bibinfo {pages} {013812} (\bibinfo {year} {2017})}\BibitemShut {NoStop}%
	\bibitem [{\citenamefont {Felicetti}\ and\ \citenamefont
		{Le~Boit\'e}(2020)}]{FelicettiPRL20}%
	\BibitemOpen
	\bibinfo {author} {S.~Felicetti}\ and\ \bibinfo {author} {A.~Le~Boit\'e},\
	\emph {\bibinfo {title} {Universal Spectral Features of Ultrastrongly Coupled
			Systems}},\ \href {\doibase 10.1103/PhysRevLett.124.040404} {\bibfield
		{journal} {\bibinfo  {journal} {Phys. Rev. Lett.}\ }\textbf {\bibinfo
			{volume} {124}},\ \bibinfo {pages} {040404} (\bibinfo {year}
		{2020})}\BibitemShut {NoStop}%
	\bibitem [{\citenamefont {Jin}\ \emph {et~al.}(2018)\citenamefont {Jin},
		\citenamefont {Biella}, \citenamefont {Viyuela}, \citenamefont {Ciuti},
		\citenamefont {Fazio},\ and\ \citenamefont {Rossini}}]{JinPRB18}%
	\BibitemOpen
	\bibinfo {author} {J.~Jin}, \bibinfo {author} {A.~Biella}, \bibinfo {author}
	{O.~Viyuela}, \bibinfo {author} {C.~Ciuti}, \bibinfo {author} {R.~Fazio},\
	and\ \bibinfo {author} {D.~Rossini},\ \emph {\bibinfo {title} {Phase diagram
			of the dissipative quantum Ising model on a square lattice}},\ \href
	{\doibase 10.1103/PhysRevB.98.241108} {\bibfield  {journal} {\bibinfo
			{journal} {Phys. Rev. B}\ }\textbf {\bibinfo {volume} {98}},\ \bibinfo
		{pages} {241108(R)} (\bibinfo {year} {2018})}\BibitemShut {NoStop}%
	\bibitem [{\citenamefont {Albert}\ \emph {et~al.}(2016)\citenamefont {Albert},
		\citenamefont {Bradlyn}, \citenamefont {Fraas},\ and\ \citenamefont
		{Jiang}}]{AlbertPRX16}%
	\BibitemOpen
	\bibinfo {author} {V.~V. Albert}, \bibinfo {author} {B.~Bradlyn}, \bibinfo
	{author} {M.~Fraas},\ and\ \bibinfo {author} {L.~Jiang},\ \emph {\bibinfo
		{title} {Geometry and Response of Lindbladians}},\ \href {\doibase
		10.1103/PhysRevX.6.041031} {\bibfield  {journal} {\bibinfo  {journal} {Phys.
				Rev. X}\ }\textbf {\bibinfo {volume} {6}},\ \bibinfo {pages} {041031}
		(\bibinfo {year} {2016})}\BibitemShut {NoStop}%
	\bibitem [{\citenamefont {Bartolo}\ \emph {et~al.}(2017)\citenamefont
		{Bartolo}, \citenamefont {Minganti}, \citenamefont {Lolli},\ and\
		\citenamefont {Ciuti}}]{BartoloEPJST17}%
	\BibitemOpen
	\bibinfo {author} {N.~Bartolo}, \bibinfo {author} {F.~Minganti}, \bibinfo
	{author} {J.~Lolli},\ and\ \bibinfo {author} {C.~Ciuti},\ \emph {\bibinfo
		{title} {Homodyne versus photon-counting quantum trajectories for dissipative
			Kerr resonators with two-photon driving}},\ \href
	{https://doi.org/10.1140/epjst/e2016-60385-8} {\bibfield  {journal} {\bibinfo
			{journal} {Eur. Phys. J. Spec. Top.}\ }\textbf {\bibinfo {volume} {226}},\
		\bibinfo {pages} {2705} (\bibinfo {year} {2017})}\BibitemShut {NoStop}%
	\bibitem [{\citenamefont {Vicentini}\ \emph {et~al.}(2018)\citenamefont
		{Vicentini}, \citenamefont {Minganti}, \citenamefont {Rota}, \citenamefont
		{Orso},\ and\ \citenamefont {Ciuti}}]{VicentiniPRA18}%
	\BibitemOpen
	\bibinfo {author} {F.~Vicentini}, \bibinfo {author} {F.~Minganti}, \bibinfo
	{author} {R.~Rota}, \bibinfo {author} {G.~Orso},\ and\ \bibinfo {author}
	{C.~Ciuti},\ \emph {\bibinfo {title} {Critical slowing down in
			driven-dissipative Bose-Hubbard lattices}},\ \href
	{https://link.aps.org/doi/10.1103/PhysRevA.97.013853} {\bibfield  {journal}
		{\bibinfo  {journal} {Phys. Rev. A}\ }\textbf {\bibinfo {volume} {97}},\
		\bibinfo {pages} {013853} (\bibinfo {year} {2018})}\BibitemShut {NoStop}%
	\bibitem [{\citenamefont {Landa}\ \emph
		{et~al.}(2020{\natexlab{b}})\citenamefont {Landa}, \citenamefont {Schir\'o},\
		and\ \citenamefont {Misguich}}]{LandaArXiv20}%
	\BibitemOpen
	\bibinfo {author} {H.~Landa}, \bibinfo {author} {M.~Schir\'o},\ and\ \bibinfo
	{author} {G.~Misguich},\ \emph {\bibinfo {title} {Correlation-induced steady
			states and limit cycles in driven dissipative quantum systems}},\ \href@noop
	{} {\  (\bibinfo {year} {2020}{\natexlab{b}})},\ \Eprint
	{http://arxiv.org/abs/arXiv:2001.05474} {arXiv:2001.05474} \BibitemShut
	{NoStop}%
	\bibitem [{\citenamefont {Yamamoto}\ and\ \citenamefont
		{Imamo\v{g}lu}(1999)}]{YamamotoBook}%
	\BibitemOpen
	\bibinfo {author} {Y.~Yamamoto}\ and\ \bibinfo {author} {A.~Imamo\v{g}lu},\
	\href@noop {} {\emph {\bibinfo {title} {Mesoscopic Quantum Optics}}}\
	(\bibinfo  {publisher} {John Wiley and Sons, New York},\ \bibinfo {year}
	{1999})\BibitemShut {NoStop}%
	\bibitem [{\citenamefont {Arkhipov}\ \emph {et~al.}(2020)\citenamefont
		{Arkhipov}, \citenamefont {Miranowicz}, \citenamefont {Minganti},\ and\
		\citenamefont {Nori}}]{ArkhipovPRA20}%
	\BibitemOpen
	\bibinfo {author} {I.~I. Arkhipov}, \bibinfo {author} {A.~Miranowicz},
	\bibinfo {author} {F.~Minganti},\ and\ \bibinfo {author} {F.~Nori},\ \emph
	{\bibinfo {title} {Quantum and semiclassical exceptional points of a linear
			system of coupled cavities with losses and gain within the Scully-Lamb laser
			theory}},\ \href {\doibase 10.1103/PhysRevA.101.013812} {\bibfield  {journal}
		{\bibinfo  {journal} {Phys. Rev. A}\ }\textbf {\bibinfo {volume} {101}},\
		\bibinfo {pages} {013812} (\bibinfo {year} {2020})}\BibitemShut {NoStop}%
	\bibitem [{\citenamefont {Arkhipov}\ \emph {et~al.}(2019)\citenamefont
		{Arkhipov}, \citenamefont {Miranowicz}, \citenamefont {Stefano},
		\citenamefont {Stassi}, \citenamefont {Savasta}, \citenamefont {Nori},\ and\
		\citenamefont {\"{O}zdemir}}]{Arkhipov2019}%
	\BibitemOpen
	\bibinfo {author} {I.~I. Arkhipov}, \bibinfo {author} {A.~Miranowicz},
	\bibinfo {author} {O.~D. Stefano}, \bibinfo {author} {R.~Stassi}, \bibinfo
	{author} {S.~Savasta}, \bibinfo {author} {F.~Nori},\ and\ \bibinfo {author}
	{{\c{S}}.~K. \"{O}zdemir},\ \emph {\bibinfo {title} {Scully-{L}amb quantum
			laser model for parity-time-symmetric whispering-gallery microcavities:
			{G}ain saturation effects and nonreciprocity}},\ \href
	{https://doi.org/10.1103/physreva.99.053806} {\bibfield  {journal} {\bibinfo
			{journal} {Phys. Rev. A}\ }\textbf {\bibinfo {volume} {99}},\ \bibinfo
		{pages} {042309} (\bibinfo {year} {2019})}\BibitemShut {NoStop}%
\end{thebibliography}
%

\end{document}